\documentclass[]{emulateapj}
\newcommand{\eg}{{\it e.g.,}~}
\newcommand{\ie}{{\it i.e.,}~}

\newcommand{\Msun}{M_{\odot}}
\newcommand{\Lsun}{L_{\odot}}

\newcommand{\LCDM}{$\Lambda$CDM~}
\newcommand{\beq}{\begin{equation}}
\newcommand{\eeq}{\end{equation}}

\newcommand{\fihl}{f_{\mathrm{IHL}}}

\begin{document}

\title{Shredded Galaxies as the Source of Diffuse Intrahalo Light On Varying Scales}

\author{
Chris W. Purcell\altaffilmark{1}, 
James S. Bullock\altaffilmark{1}, and 
Andrew R. Zentner\altaffilmark{2,3}
}

\altaffiltext{1}{
Center for Cosmology, 
Department of Physics and Astronomy, The University of California, Irvine, CA 92697 USA
}
\altaffiltext{2}{
Kavli Institute for Cosmological Physics, Department of Astronomy and Astrophysics, \& 
The Enrico Fermi Institute, The University of Chicago, Chicago, IL 60637 USA
}
\altaffiltext{3}{
National Science Foundation Fellow
}

\begin{abstract}  

We make predictions for diffuse stellar mass fractions in dark matter halos 
from the scales of small spiral galaxies to those of large galaxy clusters. 
We use an extensively-tested analytic model for subhalo infall and evolution and 
empirical constraints from galaxy survey data to set the stellar mass in each accreted subhalo, 
which is added to the diffuse light as subhalos become disrupted due to interactions 
within their hosts. We predict that the stellar mass fraction in diffuse, intrahalo light 
should rise {\em on average} from $\sim{}0.5\%$ to $\sim{}20\%$ from small galaxy 
halos ($\sim{}10^{11} \Msun$) to poor groups ($\sim{}10^{13} \Msun$). The trend with mass flattens 
considerably beyond the group scale, increasing weakly from a fraction of $\sim{}20\%$ in poor galaxy clusters ($\sim{}10^{14}  \Msun$) to roughly $\sim{}30\%$ in massive clusters ($\sim{}10^{15} \Msun$).  The mass-dependent diffuse light fraction 
is governed primarily by the empirical fact that the mass-to-light ratio in galaxy halos must 
vary as a function of halo mass. Galaxy halos have little diffuse light because they accrete most 
of their mass in small subhalos that themselves have high mass-to-light ratios; stellar halos 
around galaxies are built primarily from disrupted dwarf-irregular-type galaxies with $M_*\sim{}10^{8.5}  \Msun$. 
The diffuse light in group and cluster halos is built from satellite galaxies that form 
stars efficiently; intracluster light is dominated by material liberated from massive 
galaxies with $M_* \sim{}10^{11} \Msun$.  Our results are consistent with existing 
observations spanning the galaxy, group, and cluster scale; however, they can be tested 
more rigorously in future deep surveys. 

\end{abstract}


\maketitle

\section{Introduction} 

When Zwicky first observed the diffuse, luminous component
of the Coma cluster  of galaxies, it was  not clear what 
processes were responsible for it \citep{Zwicky51}.  
Today, the prevailing paradigm for structure formation is 
hierarchical;  galaxies and clusters of galaxies of all 
sizes are built through sequential mergers of many smaller objects.  
Hierarchical structure formation theories provide a mechanism 
for the formation of intracluster light as material lost from shredded 
galaxies over the course of cluster formation
\citep{Gallagher_Ostriker72,Merritt83,Byrd_Valtonen90,Dubinski03,
Gnedin03,Mihos04,Murante_etal04,Lin_Mohr04,Willman_etal04,
Sommer-Larsen06,Rudick_etal06,Conroy_etal07}.    Whereas the    building  blocks  of
clusters are galaxies, galaxy-sized    objects build their   masses by
acquiring    relatively  low-luminosity (dwarf)   galaxies   which may
subsequently be destroyed by   tides and heating processes to  produce
the diffuse, stellar halos around galaxies like the Milky Way
\citep{Searle_Zinn78,Johnston_etal96,Johnston_etal98,Bullock_etal01,
jsb01,bj05,Robertson_etal05,Diemand_etal05,Read_etal06,Font_etal06,Abadi_etal06}.
Whether in clusters or galaxies, we refer to  this diffuse material as
``intra-halo light'' (IHL) and adopt the symbol $\fihl$ to express the
fraction  of  the total   system  luminosity  found   in this  diffuse
component.  In this paper, we explore  the connection between the size
of a  system  and the  relative  fraction   of  its total light
contributed by intrahalo stellar material.   In particular, we predict
the mean and variance in the IHL fraction as a function of dark matter
halo mass, and  we explore the origin  of the scatter  in IHL at fixed
halo mass.

Most of  our knowledge about the IHL on galaxy scales 
($ \sim{}10^{11} - 10^{12} \Msun$) comes
from  star counts within the Local  Group.  
The stellar halo of the Milky Way contains  
$\fihl \sim{}1\%$ of  the  Galaxy's total luminosity 
\citep{Morrison_etal93,Wetterer_McGraw96,Morrison_etal00,Chiba_Beers00,Yanny_etal00,Ivezic_etal00,Siegel_etal02}.  
This number can be as large as $\fihl \sim{}2 \%$ if the 
unbound Sagittarius stream stars are included in the diffuse component 
\citep[\eg][]{Law_etal05}.

Interestingly, while the dark halo of M31 is thought to be roughly the
same size as that of the Milky Way
\citep[$M_{\mathrm{M}31} \sim{}10^{12} \Msun$, see][]{Klypin_etal02,Seigar_etal06},
the recently-discovered, metal-poor stellar halo of M31 may contain
a significantly higher fraction of that galaxy's light, $\fihl \sim{}2.5-5 \%$
\citep{Irwin05,Guhatakurta_etal05,Kalirai06,Chapman_etal06}.  If the great
Andromeda stream \citep{Ibata_etal01} were included as diffuse light, this
count would be larger.
These  observations  immediately  suggest  that   there   should be   a
substantial  spread   in IHL components  among  galaxy-sized  systems.
Detections of a stellar halo component in the  smaller disk galaxy M33
($M_{\mathrm{M}33}  \sim{}10^{11}\Msun$) have  recently been  reported
\citep{Hood_etal07,McConnachie_etal06}.    These       estimates   are
consistent with a  very  low stellar mass  fraction  in the M33  halo,
$\fihl \lesssim{}1\%$, although a higher number is not ruled out 
(A. Ferguson, private communication).

In  more distant galaxy halos, the  IHL is both   harder to detect and
more difficult to discriminate from other extended components
\citep[\eg][]{Dalcanton_Bernstein02}.
Some  results    suggest  that galactic    stellar   halos with $\fihl
\sim{}1-5\%$ are not uncommon
\citep{Sackett_etal94,Morrison_etal97,weil97,Lequeux_etal98,Abe_etal99,Zibetti_Ferguson04}.   
Recent work by \citet{buehler07} regarding the edge-on galaxy NGC 4244 
indicates the existence of an asymmetric stellar component far above the system's 
exponential thin disk, although non-detections are also reported in galaxies of similar size 
\citep[\eg][]{Zheng_etal99,Fry_etal99}.  Of particular interest is
the case of NGC 300, a low-luminosity, 
late-type galaxy in which no stellar halo has
yet   been detected,   despite   the successful  identification  of an
exponential disk  that extends over 10  scale lengths  from the disk's
center \citep{hawthorn2005}.  Of course, the differences from 
object to object  may reflect
systematic observational issues, but taken at face value, they indicate
that  the IHL fraction around galaxy halos shows significant variation
and that there may be a trend for low $\fihl$ levels
in small galaxies.  Relevant
determinations will    become more  precise  as  resolved-star surveys
extend beyond the Local Group \citep[\eg][]{2007astro.ph..2168D}.

Diffuse light fractions on  group   scales 
($M_{\rm  vir}  \sim{}10^{13} \Msun$) 
also exhibit considerable variation from system to system; however, 
the IHL component typically accounts for a more substantial fraction
of the total luminosity of the system than it does on galaxy scales.
Observations suggest that the M81 and Leo groups have at
most    a    few    percent   of their     light    in   diffuse  form
\citep{Feldmeier06,Castro-rodriguez_etal03}.  At the opposite extreme,
 HGC90 has a reported IHL 
fraction of $\fihl \sim 45 \%$ \citep{White_etal03}.  
Studies in other groups of roughly the same size
yield a range of IHL fractions, $\fihl \sim{}5 - 30 \%$ 
\citep{DaRocha_etal05,Aguerri_etal06}. 

Galaxy clusters ($\sim 10^{14} -   10^{15} \Msun$) typically show  the
highest fractions of  diffuse  light.  Again the scatter  in estimated
values is significant but values range from $\fihl \sim 10-40\%$
\citep{Thuan_etal77,Melnick_etal77,Uson_etal91,Bernstein_etal95,
cmbs00,Lin_Mohr04,Feldmeier_etal04,Mihos_etal05,Zibetti05,Krick_etal06,seigar06}.  
A review by \citet{ciardullo2004} describes 
recent surveys in the small Fornax and Virgo 
clusters and points out a distinctive fall-off in the IHL fraction
for systems smaller than $L \sim 10^{11} L_{\odot}$ -- quite similar to the break we see in
our predicted fractions below.  Another interesting, though tentative trend,
is that IHL fractions in clusters without cD galaxies appear to have a somewhat smaller
typical $\fihl (\sim 10-20\%)$, than do clusters with cD galaxies
\citep{Feldmeier_etal04,feldmeier2004b}.

Recently,         a    series        of        papers               by
\citet{Gonzalez_etal07,Gonzalez_etal05}  have   argued that   a   more
appropriate  quantity  to investigate is the   sum of the 
diffuse intracluster
light with that  of the brightest cluster galaxy 
(more generally, the ``brightest halo galaxy'' or BHG)
since  it is difficult to disentangle the two components
\citep*[the same approach is advocated by][]{Conroy_etal07}.
\citet{Gonzalez_etal05} find that the sum of IHL+BHG light
is dominated by the diffuse component on cluster scales, IHL/(IHL + BHG) $\sim 80 \%$.
Moreover,  \citet{Gonzalez_etal07} find 
that, compared to the  total light in the cluster, 
the IHL+BHG fraction {\em decreases} from  $\sim 35\%$ in 
low-mass clusters $M \sim 10^{14} \Msun$, 
to $\sim 25 \%$ in more massive clusters.
As we discuss below, these trends are very much in line with our expectations.

\begin{figure*}[t!]
\plotone{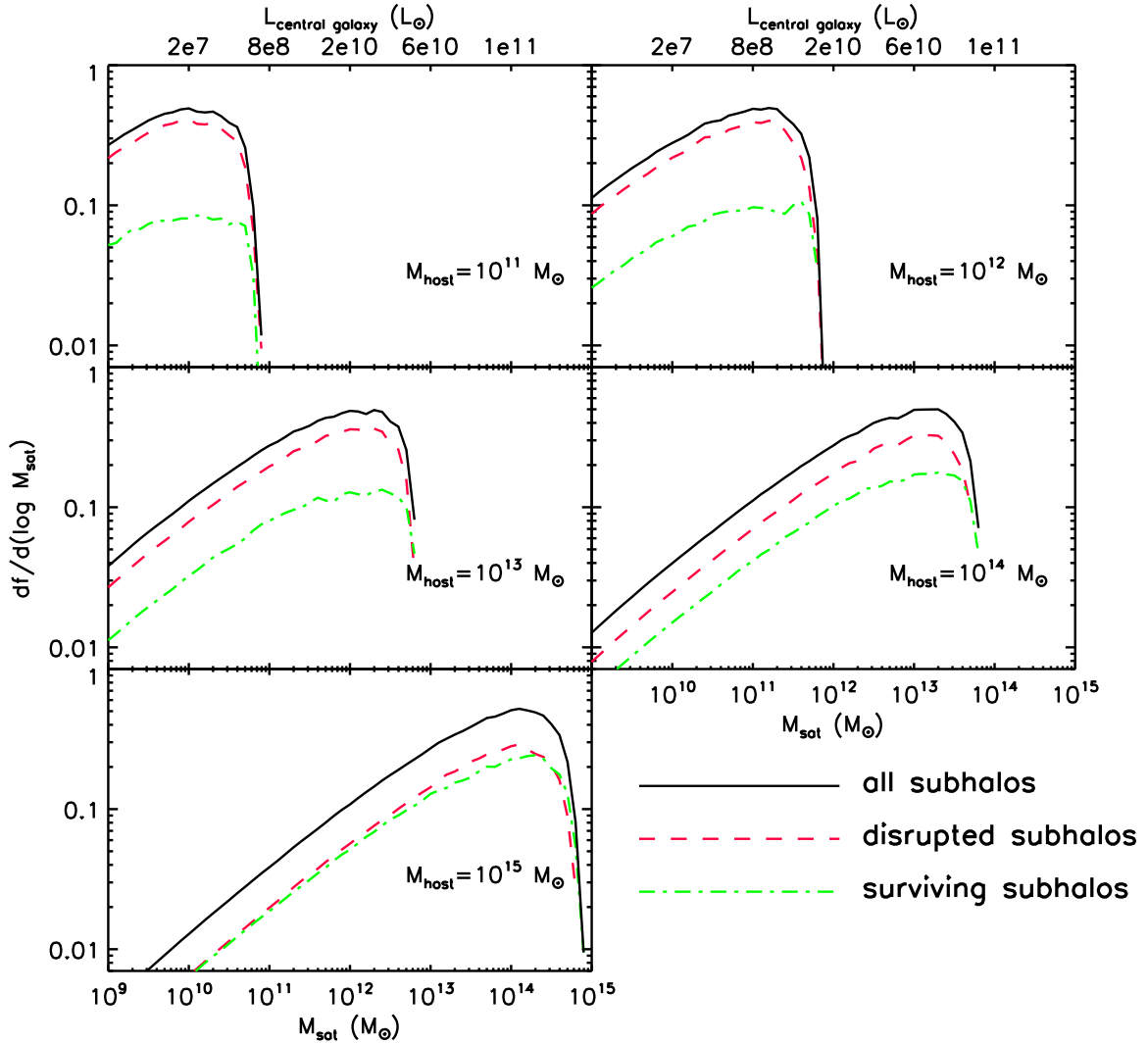}
\caption{
The differential mass fraction in subhalos of mass $M_{\mathrm{sat}}$, 
$\mathrm{d}f/\mathrm{d} (\log M_{\mathrm{sat}})$, as a function of 
satellite mass.  We plot the mass fractions for host halos of five 
masses as marked in each panel.  
The {\em solid} lines represent the mass fraction of all 
satellites accreted throughout the entire history of the host system.  
The {\em red dashed} lines represent the fraction contributed by subhalos 
that are eventually disrupted according to our algorithm, while the 
{\em green dot-dashed} lines represent halos that survive according to our algorithm.  
The upper horizontal axes show the luminosities of the galaxies assigned to 
each subhalo.
}
\label{fig:subfrac}
\end{figure*}

Comparing predictions for the IHL fraction  with observational data is
a nontrivial task.  On the  galaxy scale, total stellar halo luminosities
depend sensitively  on the difficult-to-measure
central core  radius assigned to the faint halo component.
In addition, the  IHL will
typically  have  a different color  than  the  bound light in galaxies
(because it likely traces different star formation epochs),  implying
that the IHL fraction should generally be a function of the luminosity
band or tracer populations  used to determine   it.  Moreover,  some
traditional determinations  of intracluster light have used relatively
small  patches  of sky  within the clusters  themselves, introducing a
statistical  shot-noise  error  term  into  the  inferred  IHL values.  The 
deep imaging necessary for intracluster observations is also heavily dependent 
on sky subtraction, providing another systematic barrier to precision IHL 
measurements on these scales.  Ideally, direct  comparisons between predictions and observations will
mimic the influence of particular observational techniques and choices
on theoretical predictions.    The goal of   such studies would  be to
produce predictions and observational  results that can be compared in
their detail \citep[\eg][]{Rudick_etal06,Sommer-Larsen06}.

In   this paper, our  aim is   not to  make such  detailed comparisons
between predictions and observations.   Rather, we focus on predicting
the general behavior of IHL fractions as a function of the size of the
system from dwarf      galaxies  to  large
clusters~\footnote{Note that \cite{Murante_etal04} predicted  a
positive trend between intracluster light fraction  and cluster mass, focusing 
only on cluster scales.  We  extend the range  of mass by  more
than two orders  of magnitude.}.  
We also explore the  typical galaxy size
that contributes  to IHL as  a function of halo  mass and  explore the
scatter from system to system at fixed host mass.  The scope of this study
represents a  challenge   for  direct numerical  simulation   of  halo
formation due  to the limited dynamic  range of such computations.  To
achieve our goals, we rely on  an analytic treatment of halo formation
\citep[][see below]{Zentner05}.  We normalize  the stellar  content of
our accreting halos to  match empirical constraints   from $z \sim  0$
observations \citep{yang_etal03,vale-2004-353,Bell_deJong01,deJong_Bell06}.  
We make the explicit assumption that stellar material in galaxies is liberated
when their dark matter  halos become significantly stripped.   We make
no distinction between  material that has  recently  been liberated by
tidal  interactions  (which  may  therefore  appear  as stream-like
structure)  and the general diffuse background.  In   order to avoid any
ambiguities  associated with the  evolution of luminosity in different
components,  we quote the diffuse  {\em  stellar mass} fraction,
$\fihl \equiv M_*^{\rm diff}/M_*^{\rm total}$.

In the  next   section,  we outline   our   two-step model  for    IHL  predictions.  
In \S~\ref{subsec:exp}, we briefly describe a toy model for the scaling of the 
IHL fraction with halo mass that serves both to frame our expectations for the 
fiducial result and to demonstrate the generality of 
this scaling.  In  \S~4, we present our results for  IHL
fractions, reserving \S~\ref{sec:discussion} for 
discussion and review.  Throughout this
work we  adopt    a \LCDM cosmology model  with    $h=0.7$,
$\Omega{}_{m}=1-\Omega{}_{\Lambda{}}=0.3$,  and a  primordial    power
spectrum  which    is  scale-invariant, $n=1$,   and   normalized to 
$\sigma{}_{8}=0.9$.

\section{Methods}
\label{sec:methods}

\subsection{Dark halo accretion and disruption}

We  model {\em  host} dark  matter halo  mass accretion histories  and
track  the evolution of accreted  dark matter {\em  subhalos} using an
analytic prescription developed and tested against dissipationless 
cosmological simulations by \citet[][Z05 hereafter]{Zentner05}.   
This approach is based on the earlier model of \citet{zb:03}.  
The analytic technique enables  us to explore quickly the expected
variety  of accretion  and disruption  histories for  host  halos at a
series  of different  masses.  The model has proven   remarkably  
successful  at reproducing
subhalo   count   statistics,  radial   distributions,  and  two-point
clustering  statistics measured   in  full,  high-resolution  $N$-body
simulations in regimes where  the  two techniques are   commensurable.  
This success spans more than  $3$  orders of magnitude  in  host halo mass and 
persists as a function of  redshift (Z05).  The range  over which   this
agreement is  known to exist  is limited only by  the dynamic range of
the simulations used by Z05.   In what follows,  we apply the analytic model
outside  the range over which it  is well  tested,  but we  know of no
reason that it  should fail outside  of  this range.  Of  course, more
precise estimates that   involve   full $N$-body   and  hydrodynamical
simulations will need to  be made to  refine our predictions; however,
the general success of the model  suggests that our predictions should
be  accurate enough that the  approximate dynamical treatment of
subhalos is  not  the limiting source of   error  and that  potential
differences are likely to test  our assumptions about the evolution of
stellar mass.  Even  so, many of the qualitative  trends we derive are
reflections of   very  general  features of    hierarchical  structure
formation and should be robust. 
Here we provide a brief overview of the technique and refer the reader
to \citet{Zentner05} and the similar models of 
\citet{taylor_babul04},       \citet{penarrubia_benson05},   
\citet{Faltenbacher_Mathews05}, and
\citet{vandenbosch_etal05} for more detail.

In hierarchical  cosmologies  like $\Lambda$CDM, dark  halos accumulate their
mass through  a series of  mergers with smaller objects.   The first
step  in our model  is to select a  host halo mass $M_{\rm host}$ at
$z=0$  and generate  a  subhalo-based mass accretion history  using the
extended  Press-Schechter formalism  
\citep[][for a recent review see \citealt{zentner06}]{Bond91,LC93}.  We use  the
particular implementation advocated by
\citet{Somerville99}.  The merger tree contains a list of all of
the merger times and masses of all of the smaller halos that 
have merged to form the final object.
Every time there is a merger, the smaller object becomes
a {\em subhalo} of the larger object.  This is a Monte Carlo 
procedure.  Each merger event is drawn from a probability 
distribution and by realizing merger trees for numerous 
halos of the same final mass, we can probe the variety of 
formation histories that lead to final objects of the same 
size.  As we discuss below, this variety of halo mass acquisition 
histories is a primary source of scatter in the fraction of IHL at fixed host mass.

After constructing a large number of merger histories at each final 
mass scale, we then track the  evolution of subhalos in the 
dense environments of their host systems.  
Specifically, we assign an initial  orbital energy and impact parameter 
to  each merging subhalo.  These values are chosen from 
probability distributions extracted from  cosmological 
$N$-body simulations in Z05.
We  then integrate the orbit  of  each subhalo  in  the potential of the
main halo from the time of accretion to the  epoch of observation.  We
model  tidal mass   loss  using  a  modified   tidal approximation and 
a prescription for internal heating, as well as the effect of 
dynamical friction using an adaptation of the Chandrasekhar formula
\citep{chandrasekhar43} suggested   by  \citet{hashimoto_etal03}.  For
simplicity,  we  model the density structures  of  all halos  and 
subhalos by the spherically-symmetric density profile of 
\citet[][NFW]{NFW97}.  For each halo and each subhalo, we 
set the concentration of the NFW profile according 
to the prescription of \citet{wechsler:02} to account for the 
correlation between mass accretion history and halo concentration.  Masses 
are defined relative to the virial overdensity $\Delta{}_{vir}$, where $\Delta{}_{vir}=337$ 
at $z=0$ \citep[\eg][]{bullock-2001-321}.

\begin{figure}[b!]
\plotone{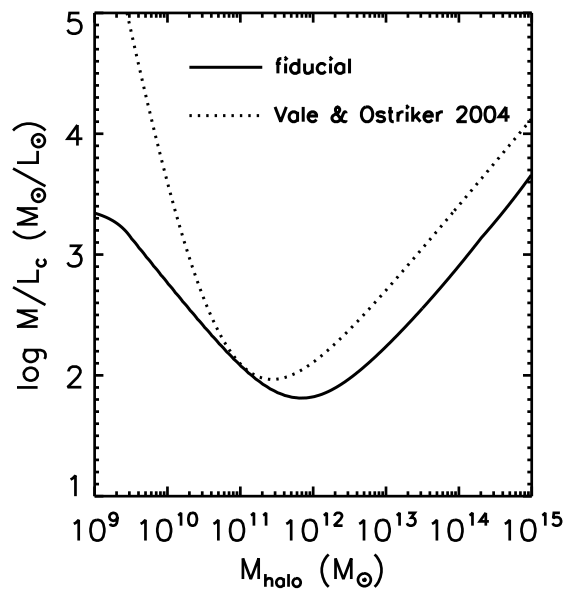}
\caption{
The total mass-to-central-galaxy-light ratio as a function of 
halo mass.  
The {\em solid} curve is the value  
inferred by \citet{yang_etal03}.  This represents the 
$L_{c}(M)$ relation that we adopt in our fiducial models.  
For comparison, the {\em dotted} line is the mass-to-light 
ratio presented by \citet{vale-2004-353}.
}
\label{fig:masslum}
\end{figure} 

Each       subhalo       has    a      well-defined    rotation  curve
$V_{\mathrm{c}}=\sqrt{GM(<r)/r}$,  that  peaks  at a velocity  $V_{\rm
max}$.  As the subhalo orbits within  its host, it gradually loses mass
at    all radii and the    value  of $V_{\mathrm{max}}$ declines.  A
subhalo
  is  declared to  be  ``disrupted''  when its maximum circular
velocity falls below  $V_{\mathrm{crit}} = f_{\mathrm{crit}} 
V_{max}(t_{acc})$.  The quantity $f_{\mathrm{crit}}$ 
is a parameter that allows us to determine when the
galaxy associated with each halo will contribute its stars to the
diffuse light of its host halo.
We have some freedom to tune $f_{\mathrm{crit}}$ 
to match
empirical constraints on the number of {\em surviving}
satellite galaxies
per halo \cite[see, \eg][and discussions below]{yang_etal03}.  
We expect that a satellite 
galaxy will typically remain bound within its subhalo
 until the subhalo loses 
a significant portion of  its mass.  Physically, $f_{crit}$
should not be so high  that a system would  be classified as disrupted
when its  host  halo is  only slightly  less  massive  than it was  at
accretion.  Similarly, a very low  choice  of $f_{crit}$ would  ensure
that  the  galaxy would  not  be  considered  destroyed until the dark
matter   in  its host   subhalo is less massive than the
galaxy itself.  

Adopting a simple mass-scaling argument may allow us to gain physical
insight into the disruption threshold, if we consider that the virial
mass  of  a halo   scales approximately as  $M\propto{}V_{max}^{3.4}$
 \citep{bullock-2001-321}.  With this  in mind, an $f_{crit}$ value of
 0.8 translates  to the halo being "disrupted"  when it  has lost just
 over  half   its   mass, while  $f_{crit}=0.2$    implies a mass-loss
 threshold of  more than 99.5\%.   Clearly, the smaller our $f_{crit}$
 is, the more  assured we can be  that galaxies meeting  the criterion
 are  truly dispersed, but if this parameter is chosen to be too small then we
may falsely associate galaxies with what should rightly be diffuse, luminous material.
As discussed below, we
adopt $f_{crit}=0.6$ as our fiducial value primarily because it
produces reasonable agreement    with  empirical constraints 
 described in  \S~\ref{subsec:light}.  This choice implies disruption 
begins to occur when just under $\sim 20 \%$ of the halo mass remains bound.

Our definition of ``disruption'' is not necessarily meant to 
indicate that beyond this threshold, a subhalo must become physically unbound 
due to the interaction within the host potential.  Rather, 
our intention is to introduce some effective criteria 
whereby it would be sensible to assign a large fraction of 
the subhalo's stellar mass to a diffuse component.  
The parameter $f_{crit}$ denotes this transition from 
a bound galaxy component that contributes little diffuse light, 
to a tenuous structure that relinquishes most of its stellar mass 
to the diffuse component of the host halo.  In our IHL predictions,  we
make  the explicit  assumption that the stars initially
assigned  to  a  subhalo   become ``diffuse''   when  that  subhalo is
``disrupted'' according to the aforementioned criterion.   

Armed with a prescription for the mass  accretion histories of halos 
and the subsequent orbital  dynamics of their satellites,  we  can
investigate  the predicted substructure distributions and overall
accretion histories for host halos  of various masses.   Our main results
rely on 1000 realizations for virial host masses
from $10^{10.5}$ to $10^{15.0} \Msun{}$, 
with four discrete intervals in each decade 
(\eg in log-space: 11.2, 11.5, 11.8, 12.0, etc.) for a total of 19 mass bins.  
The solid lines in Figure~\ref{fig:subfrac} show the fraction of host  
halo mass accreted in satellite halos
of a given mass (${\rm d} f / {\rm d} \log M_{\rm sat}$) 
averaged over 1000 realizations for host halos of
mass $M_{\rm host} = 10^{11} \Msun$ (top  left) through 
$M_{\rm host} = 10^{15} \Msun$ (bottom left)
at $z=0$.  To be explicit,  $f(>M_{\rm sat})$  is the cumulative mass fraction in
satellites larger than $M_{\rm  sat}$ and our prescription demands that
$f(>M_{\mathrm{sat}}) \to 1$ as $M_{\rm sat}  \to 0$ (\ie all of a halo's mass is
accreted in subhalos of some size).  In each panel, the dot-dashed lines include only
subhalos that survive to the present day and the dashed lines include only
subhalos that are disrupted, according to the above definition, between the 
epoch of accretion and $z=0$.  

\begin{figure}[b!]
\plotone{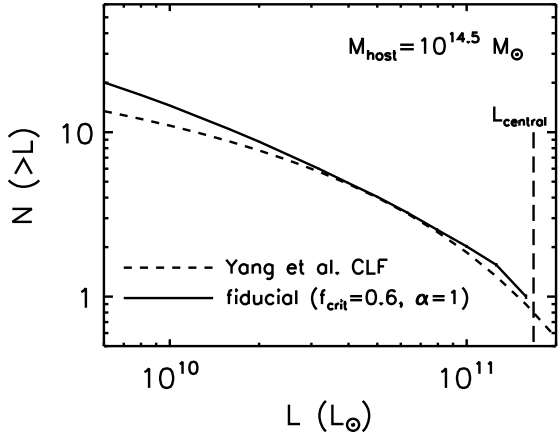}
\caption{
The cumulative number of surviving galaxies as a function of luminosity, 
both in our fiducial model ({\em solid}) and in the Yang et al. CLF analysis ({\em dashed}), 
for a host halo of mass $M_{host}=10^{14.5} \Msun$.  The model line represents the mean 
of $1000$ fiducial realizations, and the error bars representing the error on the mean over 
the sample are smaller than the model line's thickness on this plot.  
}
\label{fig:nlplot}
\end{figure}

It is important to note that regardless  of host mass, 
the majority of mass is  {\rm accreted} in  subhalos of mass 
$M_{\mathrm{sat}} \sim{} 0.05-0.1 \,  M_{\rm host}$.  In 
addition, surviving  subhalos contribute much less mass than their 
destroyed counterparts of similar size in galactic systems, 
while their relative contributions are more even in cluster-size halos.
This trend arises because high-mass 
halos accrete their subhalos more recently than
low-mass halos.  Therefore, the subhalos of low-mass halos are 
typically more dynamically evolved 
and more likely to be destroyed (see Z05).  
These facts are fundamental to understanding the 
diffuse light fractions as a function of
host mass, the consequences of which we explore in \S~3.

\subsection{Assigning Light to Dark Matter Halos}
\label{subsec:light}

We  assign a luminous  component to each accreted  halo using an
empirical model  that is normalized  to $z=0$ galaxy  constraints.  We
assume  that every  accreted subhalo  and every host  halo contains  a
central galaxy.  For  every system accreted  at time $t_{\rm acc}$  we
determine the stellar mass that this system {\em would have} today
(at $t=t_0$ or $z=0$) according to empirical mass-to-light ratios.  
Next, we extrapolate this $z=0$ value
{\em  backward} in time to obtain $M_*(t=t_{\rm acc})$
using an empirically-motivated  star formation  law.   
The $z=0$ normalization guarantees  that our model produces the
required
relationship between host halo mass and (central) galaxy luminosity
required to match 
local galaxy counts and galaxy clustering observations.

As we show below, our results for IHL fractions are quite insensitive
to star formation assumptions.    Indeed, 
our primary prediction, that the IHL fraction in halos will vary strongly
with mass scale, is extremely robust, and 
is driven by the empirical fact that the global mass-to-light ratio 
($M/L$) in \LCDM halos
must vary strongly with host halo mass
in order to reproduce the observed
 galaxy luminosity function and clustering statistics
\citep[\eg][]{White78,kwg93,sp:99,tinker:05,cm:05}.

We adopt the $M/L$ relation inferred by \citet*{yang_etal03} in their model ``M1.''  
\citet{yang_etal03} used data from the Two-Degree
Field Galaxy Redshift  Survey (2dFGRS) to constrain  the ``conditional
luminosity function'' (CLF)  of the 2dFGRS galaxies.  This  comparison
allowed them to  derive a characteristic  B-band luminosity, $L_c(M)$,
for the central (brightest) galaxies that sit in halos of virial mass $M$
\cite[for related analyses, see][]{vandenBosch_etal03,tinker:05,cm:05,yang05}.
The solid line in Figure~\ref{fig:masslum} shows  the inferred 
total mass-to-light ratio $(M/L_c)$ as a 
function of halo mass~\footnote{Data table kindly provided by X. Yang.}.  
The dotted line shows an independent result from \cite{vale-2004-353}, which we utilize 
below in order to investigate the dependence of
our conclusions on the specific choice of $(M/L_c)$ function.
Note that in both cases, galaxy formation is most efficient in
dark halos of virial mass $M \simeq{}5 \times 10^{11} \Msun$ and the conversion
of baryons to stars is increasingly
less efficient as we consider halos 
with masses either larger or smaller
than this scale.

In practice, we work with  stellar mass rather  
than   luminosity to avoid uncertainties associated with 
stellar population evolution.  
After computing the central galaxy luminosity using 
the Yang et al.~relation shown in Figure~\ref{fig:masslum}, we 
convert this luminosity to a stellar mass using the
average ``mass-dependent dust'' relation from 
\citet{Bell_deJong01}:
\beq
\left(\frac{M_*^0}{L_c}\right) = 
0.75\left(\frac{L_c}{10^{10} L_\odot}\right)^{0.33}.
\label{eq:BdJ}
\eeq
We have adjusted the \citet{Bell_deJong01}  {\em normalization}
down by a  factor of $1.26$  as advocated by their more recent
work \citep{deJong_Bell06}.
In the final analysis, our predictions for diffuse light
{\em fractions}  depend  very little on the
overall normalization.

If we were interested only in contemporary galaxy and halo properties, 
the $(M/L_c)$ relation at $z=0$ would suffice.
However, the  majority of the 
subhalos in our models are accreted well before $z=0$.  
This fact forces us to adopt a star formation prescription in
order to extrapolate our $z=0$ stellar masses to earlier times.
For simplicity, we  assume that  a  galaxy's star formation  is
truncated at the time it is accreted into a larger host, perhaps due to 
ram pressure stripping or the fact that gas leaks more readily out of the 
potential well of a subhalo located in a background host than it would 
if the satellite were left alone in the intergalactic field.  

After setting the $z=0$ stellar mass, we adopt a simple 
approach that models 
star formation with minimal parameterization, in order to estimate the stellar mass 
that a particular system would have had at the time of accretion, 
$t_{\rm acc}<t_{0}\simeq 13.6$~Gyr.   We impose a history 
\begin{equation}
M_{*}(t)=M_{*}(t_{0})\left[1-\left(\frac{t_{0} - t}{t_{0}}\right)^{\alpha}\right].
\label{eq:sf}
\end{equation}
This equation introduces a second free 
parameter  $\alpha$ into our analysis, which
can be adjusted to
produce a wide range  of evolutions for the  stellar mass in a system.
For example, $\alpha=0.25$  will cause a galaxy  to  form most  of its
stellar component within the  last two Gyr,  while  a larger value  of
$\alpha=2$  results in a system with  a  much earlier formation epoch,
increasing the lookback time to half-stellar-mass formation by roughly
a factor of  five.  
As in our choice of $f_{crit} = 0.6$ for the disruption parameter,  
we similarly adopt $\alpha = 1$ to best match 
the expected luminosity function of satellite galaxies
in host halos of a given mass from  \citet{yang_etal03}.   
We make these choices
primarily for convenience and concreteness, and we demonstrate 
in \S~4.3 that our
main results for IHL fractions are largely insensitive to these parameter choices.

An example of our (surviving) galaxy population is described by the 
cumulative luminosity function plotted in
Figure~\ref{fig:nlplot}. We caution  that this figure, unlike our main
results below, focuses on galaxy {\em luminosity} rather than stellar
mass.  While we  allow for stellar  mass buildup with  time, we do not
include any luminosity evolution,   which  should be
important for determining the B-band luminosity of cluster galaxies.  
We would expect, for example, that systems that have survived in the
cluster environment for several Gyr would have stopped forming stars
and faded in blue light.  Instead, we have used
Equation  1 to convert between  stellar mass and luminosity regardless
of the redshift at which the satellite was accreted.  
We neglect any explicit stellar population
modeling in order to keep our methods as simple as possible and to
concentrate on robust, model-independent predictions.  
We present this only
to   demonstrate the  gross  consistency  with inferred 
satellite galaxy populations in halos and do not adopt this strategy
for any of our predictions below.

The  solid line in Figure~\ref{fig:nlplot} shows   
the  cumulative number  of surviving  galaxies
(including the  central galaxy) in a  cluster-sized host halo, $M_{\rm
host} = 10^{14.5} \Msun$,  as a function   of galaxy luminosity.   The
dashed   line    shows the  empirically     derived   CLF  result   of
\citet{yang_etal03}.  Here, and for  the   rest of the
paper  unless otherwise stated,   we have  used our fiducial  parameter
choices $f_{crit}=0.6$ and $\alpha{}=1$.  Overall, the agreement
is encouraging, and we find similar results for host halos of
various masses.  We match the empirical expectation quite well
for the brightest galaxies, which is not surprising because the central galaxy is 
forced to be of the ``correct'' luminosity by construction.
We gradually begin to over-predict satellite galaxy
counts relative to the empirical line
at faint luminosities, but as we now argue, this is
not of serious concern for a number of reasons.  First, as we show below,
the vast majority of accreted stellar mass will be
contributed by the most massive accreted galaxies.  
This suggests that an accurate
reproduction of the brightest satellites is the most important
aspect of the IHL calculation.  Second, the faintest galaxies will
likely be most affected by luminosity evolution (which we do not include).
These objects tend to survive the tug of dynamical friction longer
than their more massive companions, and we expect them to
fade considerably in B-band light as
they evolve in the cluster environment.
Finally,   though  errors in   the  derived
luminosity    function     are     not  explicitly   discussed      in
\citet{yang_etal03}, the  faintest galaxies in clusters are certainly
weakly constrained by gross galaxy statistics
because they are only a minor contributor to the global
count of faint galaxies in the universe 
\citep[see, \eg the cluster luminosity functions in][]{yang05}.

\begin{figure*}[!ht]
\plottwo{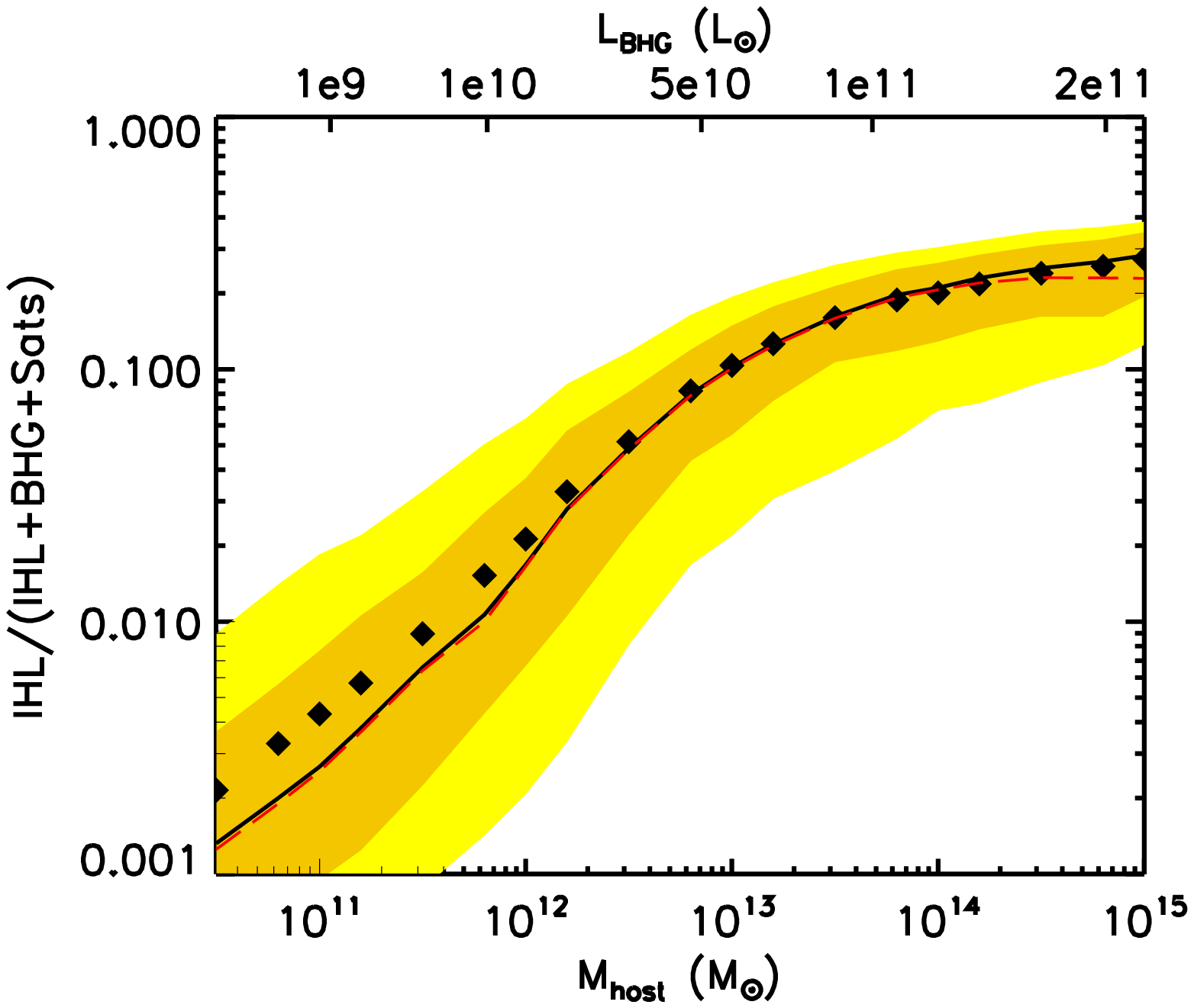}{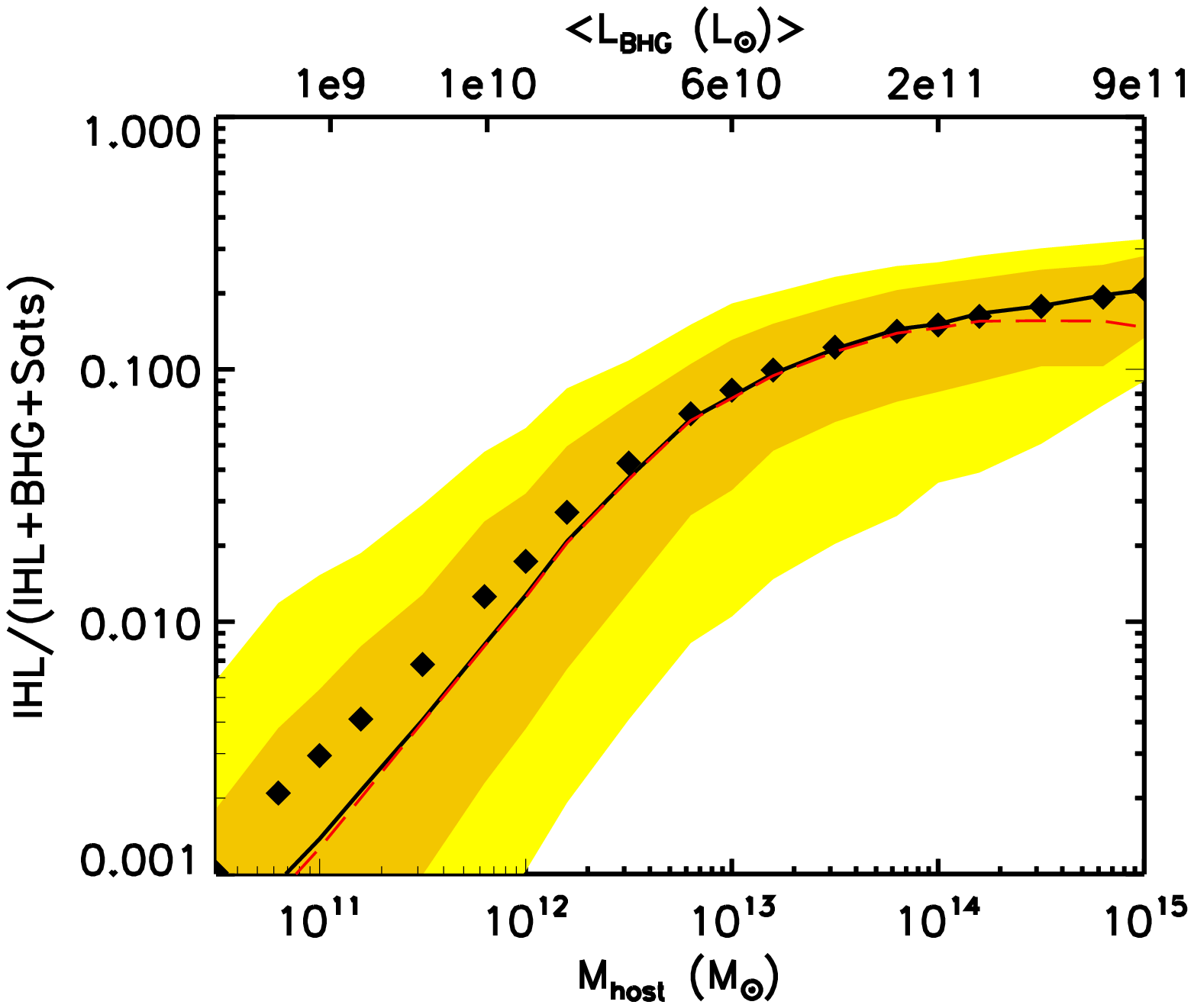}
\caption{The diffuse  light   fraction as a   function of  host halo mass,  for
systems with virial mass  between  $10^{10.5} M_{\Sun{}}$  and  $10^{15}
M_{\Sun{}}$.  In both panels the {\em diamonds} denote the mean of the distribution of 
IHL fractions at fixed mass based on 1000 realizations of our 
analytic model.  The {\em light} shaded region shows the 95\% range of 
the distribution of IHL fractions at fixed mass 
and the {\em dark} shaded region contains 68\% of  the
distribution.  The {\em solid} lines show the median of the distribution.  
Note that the median differs markedly from the mean at small host masses, 
illustrating the skewness of the IHL distribution in that range.  The {\em dashed} line 
represents the preparatory IHL fraction, without the addition of pre-processed diffuse material already in subhalos 
at the time of accretion.
The left panel shows results from Case 1 in which all stellar material from disrupted
subhalos is included as diffuse light.  The right panel shows Case 2 in 
which the subhalos on orbits passing within a critical radius $r_c$ of the host halo center contribute
their light to the central galaxy rather than the diffuse component 
(see \S~\ref{sec:methods}).  The upper axes show the corresponding central 
galaxy (BHG) luminosity derived from the prescription in \citet{yang_etal03}, with 
the Case 2 axis accounting for the average amount of stellar mass contributed by 
subhalos that merge with the BHG.
}
\label{fig:difftotal}
\end{figure*}
\subsection{Evolving the Diffuse Stellar Mass and the Central Galaxy Stellar Mass}

To  calculate the amount  of diffuse  light in a cluster, group,  or
galaxy halo, it is necessary to determine whether the stellar material
from a disrupted halo should  be included as extended, diffuse material
or as material that is incorporated into the central galaxy.  In practice, 
infalling satellites should deposit stellar mass into both the central 
galaxy and the diffuse component.  However, modeling these interactions 
in detail is challenging, so it is difficult to budget the fraction of 
the infalling stellar material that should be assigned to the diffuse 
component and the fraction that should be assigned to the central 
galaxy.  

To circumvent this complication, we employ two simple, 
alternative models for adding stellar  mass to the central galaxy
and  diffuse components  that should  bracket  the outcome from a full
modeling  of the baryonic components.   In  {\em Case  1}, we classify
{\em all} stellar  material from disrupted  subhalos as IHL.   In this
case, the diffuse stellar mass fractions  should be maximized.  
In {\em Case  2}, we exclude from  the IHL all galaxy stars from
subhalos  that  make an  approach  closer than a  radius $r_c(M_*^{\rm
host})$ to the center of their host halos.  In these instances, we add the
liberated stars to  the stellar mass of the  central object.  Relative
to Case 1,  stellar  mass is removed  from  the diffuse component  and
added to the  light of the central galaxy.   This causes diffuse
stellar mass fractions to be smaller in this case.  We   associate
$r_c$ with a  characteristic outer radius for the  central galaxy.  To
be  conservative, we  adopt a   fairly large   outer radius  
$r_{c}  = 10$~kpc, for central  galaxies of stellar  mass $\hat{M_*} = 4  \times
10^{10} \Msun$.  We assume that $r_c$ scales according to the findings
of \citet{Shen_etal03} for Petrosian half-light radii of 
galaxies in the Sloan Digital Sky  Survey.  
To be explicit, we  use 
$r_c \propto  M_*^{0.4}$  for $M_* > \hat{M_*}$  and
$r_c \propto  M_*^{0.16}$ otherwise.  The generous value of 
$r_c$ along with the assumption that all stellar mass is assigned to 
the central galaxy and that none of the stellar mass goes into the 
diffuse component should lead to minimum IHL fractions in Case 2.

Thus far, we have only considered the disruption of subhalos belonging 
to the trunk level of the host halo's merger tree, \ie we have not made any 
determinations about the diffuse stellar content already present in accreting 
subhalos, often referred to as "pre-processed" intrahalo 
light \citep[see, \eg][]{Rudick_etal06}.  We do not expect galactic-scale 
host halos to carry much of this pre-processed stellar material, since 
accreting dwarf satellite galaxies typically have very little of their luminosity 
in diffuse form, but cluster-sized hosts accrete most of their mass in galaxy 
groups that may have a significant amount of IHL already present.  In order 
to replicate this phenomenon, we first obtain our fiducial result (without 
the presence of pre-processed IHL), which is then used to interpolate 
an initial IHL value for each accreting subhalo.  We then reproduce the fiducial 
IHL fraction, this time including diffuse stellar mass {\em already present} in subhalos 
and contributing that amount to the host's total IHL upon the subhalo's accretion, 
essentially bootstrapping Case 1 into itself in order to account for 
pre-processed intrahalo luminosity.  We expect this model to differentiate itself 
from the initial result on large mass scales, at which subhalos are likely to have 
diffuse stellar components that contribute a non-negligible portion of the total subhalo 
luminosity.  It is also worth noting that this second-order substructure is likely to 
carry increasingly less IHL than the fiducial model predicts at fixed mass because these subhalos 
will be younger and less luminous and will be less dynamically evolved so less luminosity could 
have been converted to diffuse luminosity.  This 
indicates that our pre-processing method will slightly overestimate the contribution 
to the total host IHL made by diffuse stellar mass belonging to higher-order subhalos.
\begin{figure*}[!ht]
\plottwo{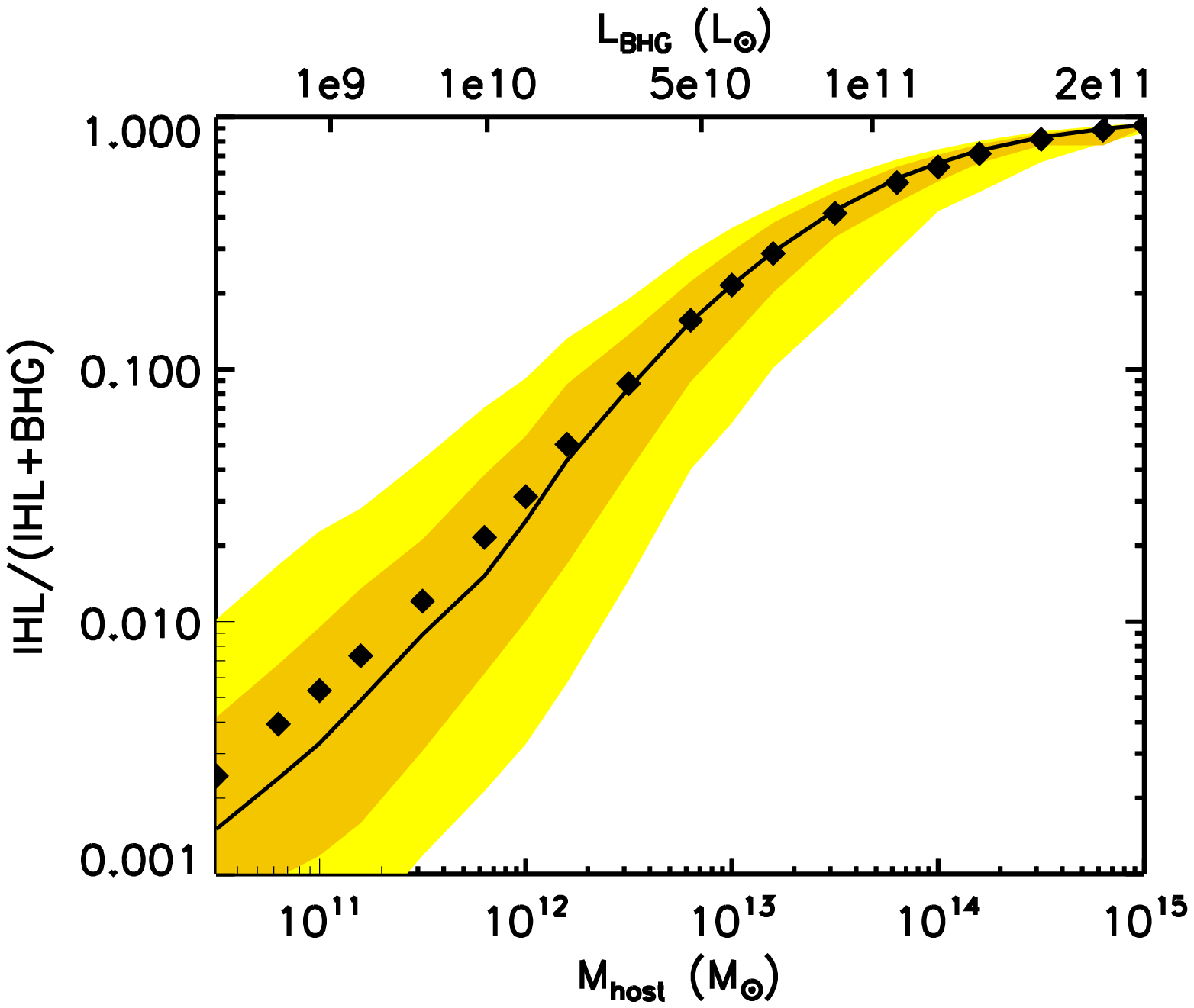}{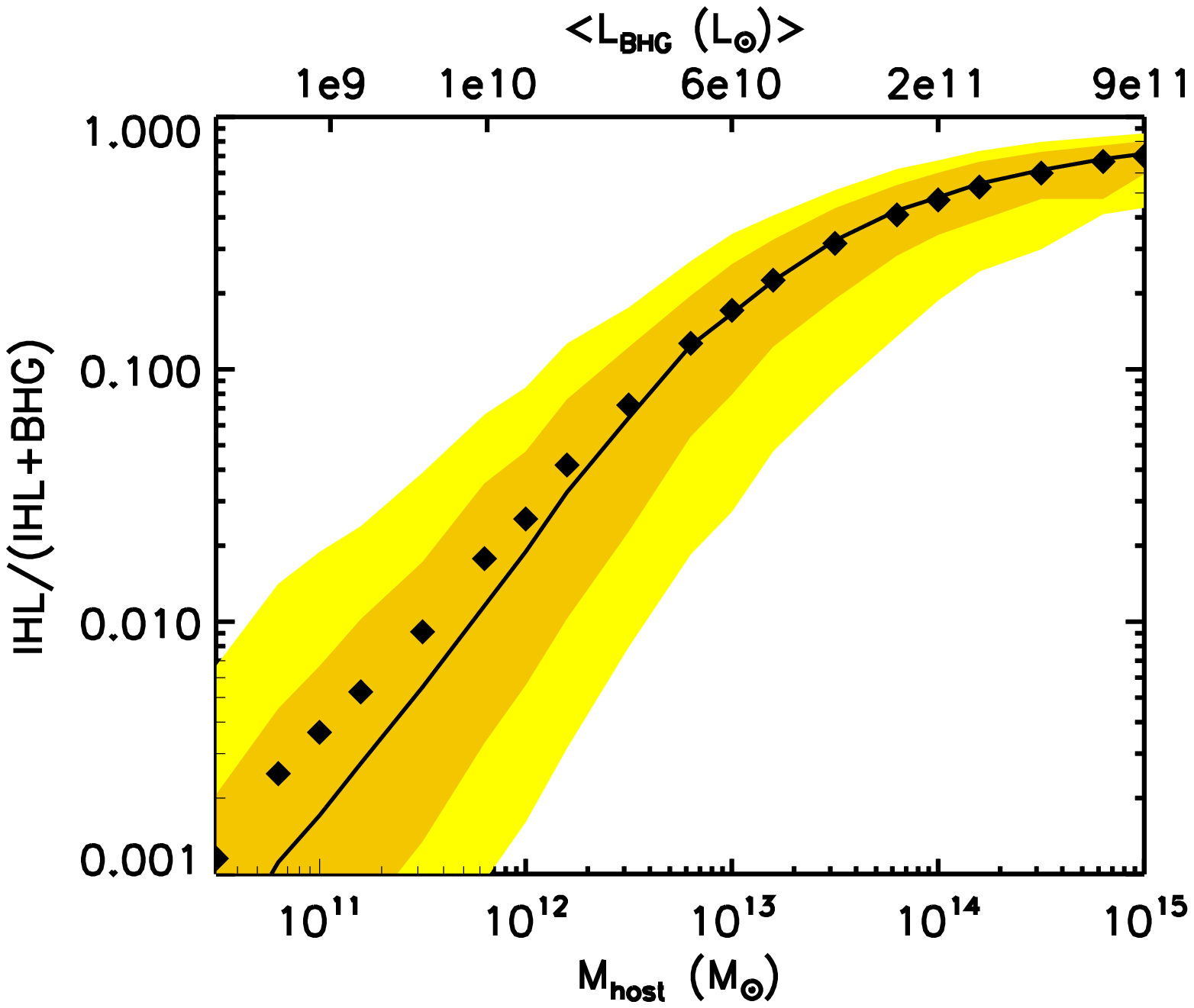}
\caption{Similar to Figure~\ref{fig:difftotal}, where now we
plot the IHL fraction relative to the sum of the IHL and
the brightest halo  galaxy (BHG) .  The rise  at high host masses
compared to Figure~\ref{fig:difftotal}  reflects  the 
fact that much of the stellar mass in clusters is bound to satellite galaxies.
Note that the IHL mass dominates that of the central galaxy mass on cluster
scales, in accord with observations \citep{Gonzalez_etal05}.  }
\label{fig:difftotal2}
\end{figure*}

\section{A Toy Model for the Intrahalo Light Fraction}
\label{subsec:exp}

Before proceeding, we derive a crude, analytic estimate for the 
scaling of the IHL fraction as a function of host halo mass.    
This model serves to frame our expectations for the general 
behavior of IHL fraction with mass, to highlight the  
features of hierarchical structure formation models most 
relevant to the determination of IHL fractions as a function 
of halo mass, and to demonstrate the generality of the 
halo mass-IHL fraction 
trends that we present in more detail in the following section.

The gross scaling of IHL fraction with halo mass  
can be understood from two robust, cosmologically-motivated inputs:
\begin{enumerate}
\item Host halos of mass $M$ tend to
accrete most of their  mass in  subhalos of mass  $M_{\rm
sat} \sim{}0.05-0.1 \, M_{\rm host}$ (Figure~\ref{fig:subfrac}), and these 
halos are disrupted very efficiently due to dynamical friction.
\item Galaxy  formation picks out a typical
halo mass  $M_t \simeq 5 \times 10^{11}   \Msun$, where  star  formation is  most
efficient, and the efficiency of star formation declines rapidly away from this value 
(Figure \ref{fig:masslum}).   
\end{enumerate}

To begin with, it is useful to introduce an
approximate analytic fit to the adopted 
$(M/L_c)$ relation from Yang et al. (2003):
\begin{equation}
\frac{M}{L_c(M)} \simeq 50 \left[\frac{M}{M_t}\right]^{-3/4}
 \left[1 + \left(\frac{M}{M_t}\right)^{3/2}\right].
\end{equation}
The differential contribution to the IHL fraction  from a satellite of
mass $M_{\mathrm{sat}}$ can be computed by introducing two parameters:
$f_{\mathrm{destroy}}$  that  encapsulates  the probability  that this
satellite will be destroyed,  and $f_{\mathrm{diff}}$  which describes
the  fraction of the satellite's  stellar mass that contributes to the
diffuse light once  it is destroyed.  Conceptually, this decomposition
is useful, because $f_{\mathrm{destroy}}$ has  a known dependence upon
host and satellite halo masses (Z05,  see Fig.~\ref{fig:subfrac}).  We
will show that this dependence is subdominant,  so for our purposes we
can  condense  these into     a single parameter   $f_{\mathrm{d}}   =
f_{\mathrm{destroy}}\ f_{\mathrm{diff}}$ that accounts for the average
fraction of its total stellar mass that a satellite contributes to the
IHL.      As    we   stated  above,    the     mass   dependence    of
$f_{\mathrm{destroy}}$ is  weak and is  not  the dominant factor  that
gives  rise   to  the  mass  scaling of   the   IHL fraction  and, for
simplicity, we will assume the composite parameter $f_{\mathrm{d}}$ to
be a slowly-varying function of mass.

The differential contribution to the IHL fraction from satellite halos in the 
mass range $\mathrm{d}M_{\mathrm{sat}}$ around $M_{\mathrm{sat}}$ is then 
\begin{eqnarray}
\frac{\mathrm{d}\fihl}{\mathrm{d}M_{\mathrm{sat}}} & = & 
f_{\mathrm{d}}\ \frac{L(M_{\mathrm{sat}})}{L(M_{\mathrm{host}})}\ 
\frac{\mathrm{d}n_{\mathrm{acc}}}{\mathrm{d}M_{\mathrm{sat}}} \\
 & \sim & 
f_{\mathrm{d}}\ \bigg(\frac{M_{\mathrm{sat}}}{M_{\mathrm{host}}}\Bigg)^{3/4}\ 
\frac{1 + (M_{\mathrm{host}}/M_t)^{3/2}}{1 + (M_{\mathrm{sat}}/M_t)^{3/2}}\ 
\frac{\mathrm{d}n_{\mathrm{acc}}}{\mathrm{d}M_{\mathrm{sat}}}, 
\end{eqnarray}
where $\mathrm{d}n_{\mathrm{acc}}/\mathrm{d}M_{\mathrm{sat}}$ is the mass 
function of accreted satellites.  In general the amount of total and 
stellar mass accreted into the system is dominated 
by the few most massive satellites near $\sim M_{\mathrm{host}}/20$ (see 
Fig.~\ref{fig:subfrac}).  As a final rough approximation, 
we assume that satellites of this mass dominate the integral over $M_{\mathrm{sat}}$.
This gives 
\begin{eqnarray}
\fihl(M) & \sim & f_{\mathrm{d}} n_{\mathrm{eff}} 
\frac{L(M_{\mathrm{host}}/20)}{L(M_{\mathrm{host}})} \\
& \sim & 
0.005 f_{\mathrm{d}}\ n_{\mathrm{eff}}\ 
\frac{ \left[\, 1 + \left(M_{\mathrm{host}}/M_t\right)^{3/2} \, \right]}
{\left[1 + \left(M_{\mathrm{host}}/20M_t\right)^{3/2}\right]}, 
\end{eqnarray}
where we have introduced a final parameter $n_{\mathrm{eff}}$ which 
represents an effective number of satellites near mass 
$M_{\mathrm{sat}}=M_{\mathrm{host}}/20$ 
and will be of order unity (Fig.~\ref{fig:subfrac}) and 
$f_{\mathrm{d}}$ is understood to be evaluated near 
$M_{\mathrm{sat}}=M_{\mathrm{host}}/20$.  

As will be clear in the following section, 
this extremely simple model captures the general features of 
our more detailed predictions.  In our full model, 
$f_{\mathrm{d}}$ should be less than one and $n_{\mathrm{eff}}$ should 
be of order unity.  This simple model predicts that 
the IHL fraction should have a small and nearly constant value 
below $M_t \sim 5 \times 10^{11} \Msun$, $\fihl \sim 5 \times 10^{-3}$.  
We expect a rapid rise in the IHL fraction with halo mass, 
 $\fihl \propto  M^{3/2}$, for 
halos in the mass range $M_t \lesssim M_{\mathrm{host}} \lesssim 20 M_t$.  
In physical units this range is 
$5 \times 10^{11} \Msun \lesssim{}M_{\mathrm{host}} \lesssim{}10^{13} \Msun$, and represents 
the range of transition between MW-like galaxies and small groups of galaxies.  
For host halos more massive than groups, 
$M_{\mathrm{host}} \gtrsim{} 20 M_t \simeq{} 10^{13} \Msun$,
both the relevant satellite halos and host halos fall along the
power-law regime of the $M/L_c$ function and
we expect the IHL fraction to remain roughly constant, 
$\fihl \lesssim 40 \%$.

At this point, it behooves us to summarize the points 
that this model illuminates regarding the IHL on different 
scales.  In our model, it is approximately true that only the 
relative sizes of host and satellite objects determine 
the probability for satellites to deposit their stellar 
mass into the diffuse component.  
Halos acquire most of their mass, dark or stellar, 
in a relatively small number of 
accreting objects of order 1/20 the size of the parent 
object (see Fig.~\ref{fig:subfrac} and Fig.~\ref{fig:dfdmstar} in the following section).  
Though the details are not known, it is an empirical fact 
that in a hierarchical cold dark matter cosmology, the 
process of galaxy formation must pick out a halo mass 
scale where galaxy formation is most efficient 
($M_t \sim 5 \times 10^{11} \Msun$) and that this efficiency 
drops at both lower and higher masses.  Halos less massive 
than $\sim 20M_t$ will accrete little stellar mass in 
satellite objects and thus have little opportunity to 
build a diffuse, stellar halo.  Halos more massive 
than $\sim 20M_t$ will accrete many satellite halos with masses such 
that they form stars at near peak efficiency.  As these host halos bring in satellites with 
lots of stars, they have ample opportunity to build diffuse stellar 
halos.  The general conclusion that diffuse light fractions should increase 
from very small values in galaxy-sized systems to larger values in 
group- to cluster-sized systems seems difficult to avoid in the 
context of hierarchical cold dark matter structure formation.

 \section{Results}
\label{sec:results}

\subsection{IHL Fraction and Dark Halo Mass}
\label{sub:ihlresults}

The two panels  of  Figure~\ref{fig:difftotal} show our 
primary results.  The predicted diffuse stellar mass 
fraction, $\fihl \equiv M_*^{\rm diff}/M_*^{\rm total}$, 
is shown as a function of host halo mass.  The total 
stellar mass $M_*^{\mathrm{total}}$ includes the stellar 
mass in the diffuse component (IHL), satellite galaxies, and
the central galaxy.  In this section we will refer to the
central galaxy as the 
``brightest halo galaxy'' or BHG, in analogy with 
the brightest cluster galaxy (BCG) in clusters.
The left panel shows  results for Case 1 in which 
we  assign all  light from  disrupted  subhalos to  the  diffuse
component.  The right panel shows Case 2 in which we 
exclude any stars that were in subhalos having passed 
within $r_c$ of the host halo center from the diffuse 
component and instead add this material to the stellar 
mass of the central BHG.   
 Diamonds show the average value of 
$\fihl$ and the thin solid line shows the median.  
These results are derived from 1000 realizations for each
host halo mass.  The light and dark shaded regions span the 95\% and
68\% regions of the distribution respectively, centered on the median.

\begin{figure}[!b]
\plotone{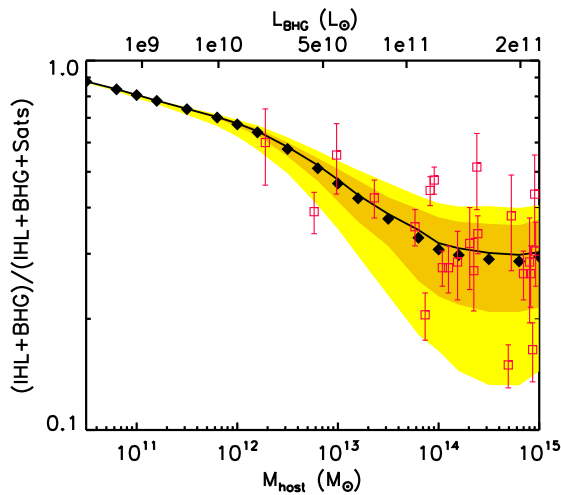}
\caption{Again similar to Figure~\ref{fig:difftotal}, 
but now plotting the IHL+BHG fraction relative to the total stellar mass
in each halo.  The open symbols with error bars are observational points
from \citet{Gonzalez_etal07}.  The predicted and observed trends show
remarkable agreement.  Case 1 and Case 2 are identical in this plot as a function of $M_{\rm host}$.
We use Case 1 to set the $L_{\rm BHG}$ values across the top axis.}
\label{fig:difftotal3}
\end{figure}

The upper axis in the Case 1 (left) panel of Figure~\ref{fig:difftotal}
shows the luminosity of the 
central galaxy according to the \citet{yang_etal03} 
mapping, while the Case 2 (right) panel upper axis label gives the
mean BHG luminosity at a given host mass that we obtained
by averaging the total merged 
subhalo luminosity (plus the assigned central galaxy luminosity).
We should note here that our Case 2 model is not self-consistent, in that we first assign 
a central galaxy luminosity according to the $z=0$ conditional luminosity function and then 
subsequently add stellar mass via subhalo mergers, which will obviously produce incorrect 
present-day stellar mass functions.  However, on cluster scales, the Case 2 
central galaxy luminosities are many times larger than that required by the \citet{yang_etal03} analysis
(\eg $L_{\rm BHG} \sim 9 \times 10^{11} L_\odot$  
compared to $\sim 2 \times 10^{11} L_\odot$ from Yang et al.), implying that even if 
our BHG were composed entirely of merged material (without {\em any} stellar mass produced by direct 
cooling processes), we would still over-predict the luminosity of the central object.  We therefore present the 
Case 2 analysis only as a means of minimizing intrahalo light production by 
dynamical considerations alone.  We note that the result of this investigation, despite the above caveat, 
is only a systematically mild reduction in IHL across the full spectrum of host mass.   
Our rejection of Case 2 aligns with the findings of \citet{Conroy_etal07}, in which the authors use a 
numerically-motivated model for the construction of massive galaxies and find that the large majority 
of centrally-merging stellar mass ($\gtrsim{}80 \%$) must be ejected into the intracluster medium 
in order to reproduce the observed evolution of these central galaxies at low redshift.

The  most  obvious trend in Figure~\ref{fig:difftotal}  is  that the 
IHL fraction rises with halo mass from galaxy to group mass scales.  {\em On
average} the  diffuse fraction  is predicted to be negligible in  
$M \lesssim{}10^{11} \Msun$  halos and quite  substantial   in groups and
clusters.   This is independent of  the method used to assign stripped
stellar  mass to the  diffuse component  or  the central  object. 
In  both cases, the relation  between $\fihl$ and  halo mass
flattens considerably at masses above the group  scale, tending towards 
a weaker evolution from a diffuse stellar mass fraction of 
about $\fihl \sim{}20 \%$ at a host mass of $\sim{}10^{14} \Msun$, to 
a value of nearly $\fihl \sim{}30 \%$ at $M_{\rm host} \sim{} 10^{15} \Msun$.  
We also see from the figure that the initial IHL fraction (without the inclusion 
of pre-processed diffuse stellar material) is virtually flat on cluster scales, 
implying that the {\em a priori} presence of subhalo IHL is largely responsible for 
the weak increase in the host's total diffuse light on those scales.  This mild trend 
has also been recovered by numerical simulations \citep{Murante_etal04,Monaco_etal06,murante07} 
as well as observations of intracluster luminosity as a function of cluster richness \citep[\eg][]{Zibetti05}.

\begin{figure*}[t!]
\plotone{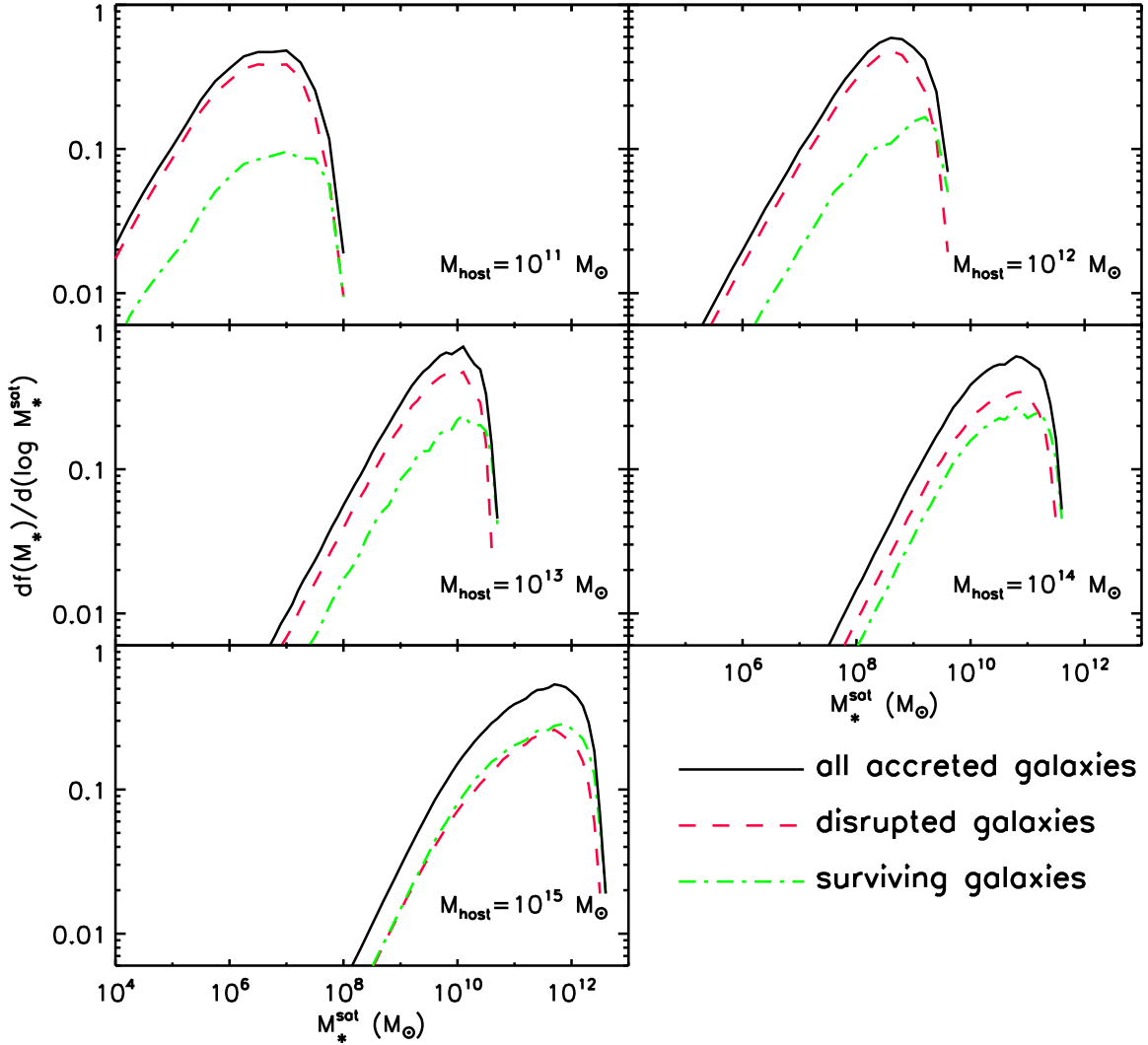}
\caption{
Similar to Figure~\ref{fig:subfrac}, the differential contribution to the {\em stellar} 
mass from subhalos with an initial stellar 
mass $M_{*}$ as a function of $M_{*}$.  We show this differential contribution 
for five host halo masses indicated in the legends of the figure.  
{\em Red dashed lines} 
indicate stellar material contributed by disrupted satellite galaxies, while 
{\em green dot-dashed lines} indicate the fraction of stellar mass in surviving subhalos for a 
particular host.
}
\label{fig:dfdmstar}
\end{figure*} 

In accordance with our simple model of the previous section, 
the trend of increasing IHL with halo mass is set primarily by  the
convolution of the distribution of subhalos that are disrupted 
(Figure~\ref{fig:subfrac}) 
with the mass-to-light ratios of halos (Figure~\ref{fig:masslum}).  
Consequently, the trends predicted by our full model follow closely our general expectations 
described in \S~\ref{subsec:exp}.  Specifically,  galaxy halos
with $M \simeq{}M_t \simeq{}5  \times 10^{11} \Msun$  
have massive central galaxies because they 
sit in the valley of the $M/L_c$ curve; however, these galaxies 
have low diffuse light contributions because
they accrete and destroy most of their mass in subhalos
of mass $M \simeq 2.5 \times 10^{10} \Msun$, where star formation is inefficient.
Halos at the group  scale ($\sim10^{13}  \Msun$) accrete large numbers  of
subhalos  near the   valley  of the   mass-to-light   ratio curve.  These
accreted satellites are a copious source  of stellar material for
diffuse light in groups.  The diffuse light fraction
begins to flatten above the group scale because both the
host {\em and} destroyed subhalos  have masses
$M \gtrsim M_t$, which corresponds to a regime where the $M/L_c$ relation
follows an approximate power law.  In this case, the ratio of destroyed satellite
luminosity to central host luminosity is independent of
mass, $L_c(M_{\rm sat})/L_c(M_{\rm host}) \sim {\rm constant}$.
We note here that there is also a 
subdominant effect that contributes to the flattening of $\fihl$ 
at high masses, namely that more massive host systems 
typically accrete their material more recently.  This leaves
relatively little time to disrupt satellites (see Z05) and results in
a lower fraction of diffuse, stripped material.

In order to explore how the total stellar mass within halos is divided among
the various components (IHL, BHGs, satellites) and to more directly compare 
our results with the variety of observational estimates in the literature, 
Figures~\ref{fig:difftotal2} and \ref{fig:difftotal3} show two alternative quantities.
In Figure~\ref{fig:difftotal2}, we ignore surviving 
satellite galaxies altogether in order to determine the relative importance of IHL as 
compared to the total stellar mass in the BHG+IHL.  In Case 1 (left) the 
IHL dominates the BHG on cluster scales, contributing $80-90\%$ 
of the combined stellar mass, while the fraction IHL/(IHL+BHG) declines to 
$\sim 1\%$ on galaxies scales, where it is nearly identical to our definition 
of the intrahalo light fraction.  Again, this is easy to understand in terms of 
the empirically-determined mass-to-light ratios of halos in hierarchical cold dark matter cosmologies.  
On galaxy scales, the host halo forms stars at near maximal efficiency, while its accreted 
substructures carry comparatively little stellar mass.  The IHL/(IHL+BHG) fraction is nearly 
equal to $\fihl$ because nearly all of the luminosity in non-diffuse (or, for that matter, diffuse) 
form is in the BHG.  As host halo mass increases, the efficiency of 
galaxy formation in the central system itself declines, meaning relatively more of the 
non-diffuse light is carried by the satellites that are not shredded.  This causes 
the IHL/(IHL+BHG) fraction to increase more rapidly with mass than $\fihl$.  
Importantly, our result compares favorably to the $\sim 80 \%$ IHL to IHL+BHG
fraction found by  \citet{Gonzalez_etal05} in galaxy clusters.  

\begin{figure*}[t!]
\plottwo{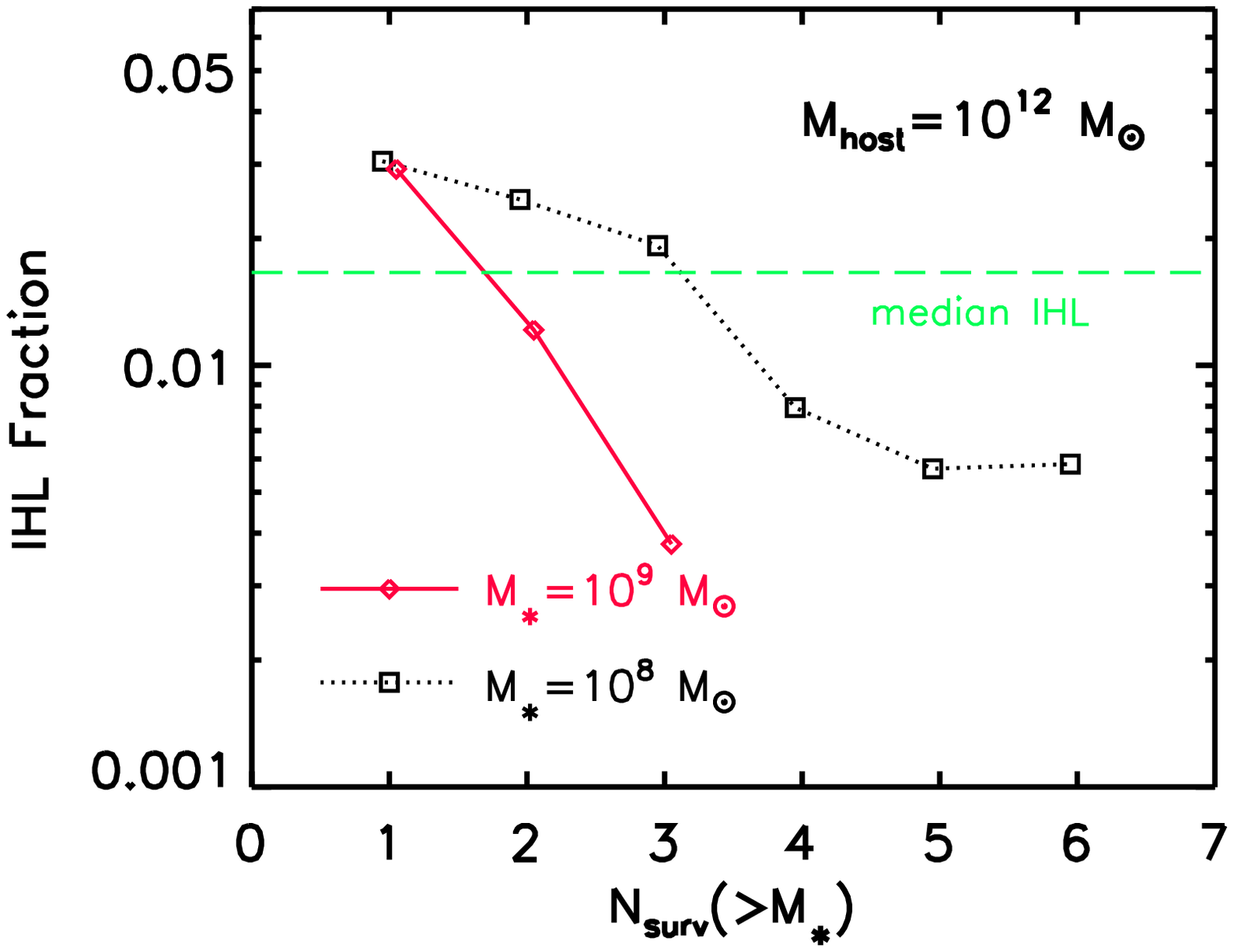}{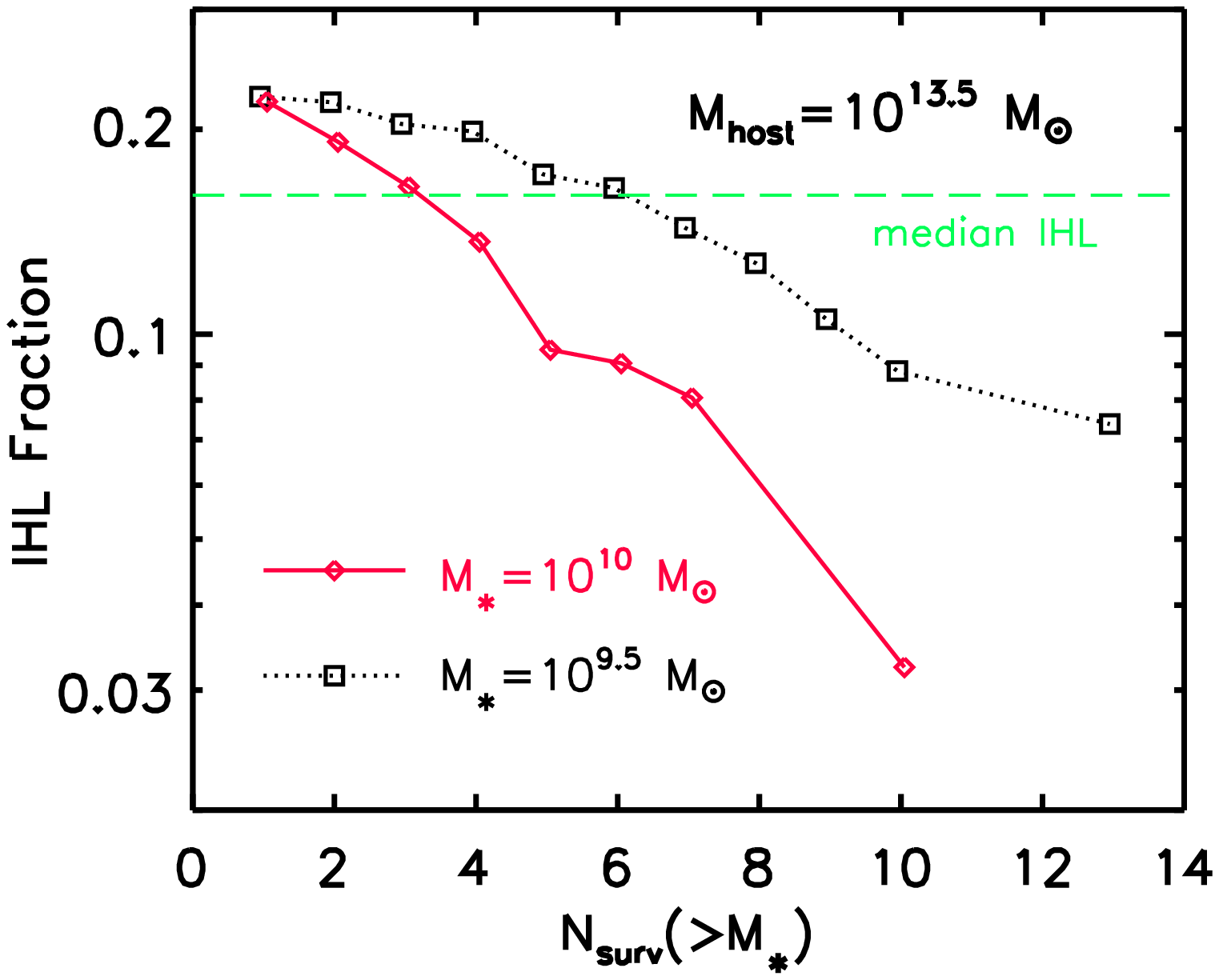}
\caption{
The median diffuse 
light fraction (Case 1) at {\em fixed host mass} 
as a function of the number of massive satellites surviving over
the halo's history.  
In the left panel we focus on $10^{12} \Msun$ halos (median $\fihl \sim 1.7\%$)and look at the
number of surviving subhalos with more stellar mass than $M_* = 10^{9}$ (diamonds)
and $M_* = 10^8 \Msun$ (squares).  In the right panel we consider
a more massive, $10^{13.5} \Msun$ host halo and show the 
diffuse light fraction (median $\fihl \sim 16\%$)
as a function of surviving subhalos with more stellar mass than
$M_* = 10^{10} \Msun$ (diamonds) and $M_* = 10^{9.5}$ (squares).  Though not shown here, the 68\% 
scatter about the median for the $10^{12} \Msun$ host is roughly constant at $\sim{}\pm{}0.3$ in 
log$_{10}$, while the $10^{13.5} \Msun$ host exhibits a smaller variance of $\sim{}\pm{}0.1-0.2$ in log$_{10}$ that 
grows slightly as $N_{surv}$ increases.
}
\label{fig:nsurv}
\end{figure*} 

Figure~\ref{fig:difftotal3} depicts a related quantity, the IHL + BHG fraction
relative to the total stellar mass.
The IHL + BHG fraction is anti-correlated  with host mass,  decreasing from $\sim 40\%$ 
on group scales to $\sim 30\%$ within large clusters.  The trend follows from 
the same logic used in the previous paragraph.  In addition to the evolution of 
$\fihl$ with mass, the BHG becomes increasingly less luminous relative to the 
sum of the luminosities of its satellite galaxies as halo mass increases.  
The open points with error
bars show the same quantity derived observationally for individual clusters and groups by
\citet{Gonzalez_etal07}~\footnote{The data table was kindly provided by A. Gonzalez.
It included information on   halo masses, $M_{500}$, within  a
radius,  $R_{500}$, where the overdensity is $500$.  The table also included
measured IHL + BHG
fractions  within $R_{500}$ and within $R_{200}$ -- the radius corresponding to an overdensity
of $200$.  For
the points plotted   on Figure~\ref{fig:difftotal3}, we  converted the
reported masses to our $\Delta_{vir} = 337$ convention for virial mass
and  plotted the  average of the IHL + BHG fractions within  $R_{500}$ and 
within $R_{200}$.  The error  bars  reflect   the larger   of the  two  reported
measurements. These corrections amounted to $\sim 20 \%$ and $\sim 5 \%$ changes in
mass and (IHL + BHG) fractions, respectively, and do not affect the overall trends
in any significant way.}.  The  predicted  and  observed  trends  are  remarkably
consistent, especially on average.  Given the observational
uncertainties, the variance in the observed points
at fixed mass is also consistent with our prediction, however
there is a tendency for the data points to
skew into the upper range of our model's scatter.  This may reflect a
bias in the observational sample, which is selected to include
systems with dominant
BHGs.  Indeed, a positive trend between
dominance of the central BHG and IHL fraction is
seen in our models (see \S~\ref{sub:distribution}).

Figure \ref{fig:dfdmstar} shows the average fraction of diffuse light that comes from 
satellite  galaxies of a given stellar mass $M_*$, 
for several choices of host dark matter halo mass.
We see that the diffuse component (or stellar halo) around small $M_{\rm host}\sim10^{11} \Msun$ (\eg M33) 
dark matter halos is built up from disrupted dwarf spheroidal-type 
galaxies with $M_* \sim 10^6 \Msun$. Stellar halos around
larger Milky-Way-type galaxies, $M_{\rm host}\sim10^{12} \Msun$, 
are built from dwarf-irregular-size systems, $M_*\sim10^{8.5} \Msun$, and 
intracluster light is produced by massive galaxies, 
$M_*\sim10^{11} \Msun$ 
\citep[see][for a similar result from numerical simulations of intracluster stars]{murante07}.
This fact is likely to be an important ingredient in 
understanding the metallicities of diffuse stellar components
as a function of galaxy luminosity \citep{2005ApJ...633..821M,ferguson07}; 
specifically, more luminous galaxies
are expected to be surrounded by more metal-rich stellar halos
because their halos are formed from more massive satellites.
Additionally, note that the differential stellar mass distributions 
become more sharply peaked as host halo mass increases from galaxies to groups, 
reflecting the increase in relative subhalo luminosity as we approach the 
$M\sim{}M_t$ valley in the $M/L_c$ relation
(Figure~\ref{fig:masslum}).  Correspondingly, the distributions broaden once
more as we consider the most massive hosts because their subhalo populations have 
moved in large part to the right of the valley.  

\subsection{The Distribution of the Diffuse Light Fraction at Fixed Halo Mass}
\label{sub:distribution}

A second important feature of the diffuse stellar mass fraction is the
relative scatter  at fixed mass,  particularly in low-mass halos.  The
width of the  distribution is driven  primarily by differences in mass
accretion histories of objects of fixed final halo mass, including the
stochastically-driven  properties of the  host's recent  merger events
and the particular orbital parameters for each plunging satellite.  As
a general rule, we expect halos that acquired their mass more recently
to have had relatively less time to disrupt the subhalos they host and
to have  less IHL, while  early-forming  host halos will  display  the
opposite behavior.  Continuing with   this logic, the number of  bound
satellite galaxies should anti-correlate    with the IHL   fraction in
objects.  Indeed,  Figure~\ref{fig:nsurv}  illustrates that  our model
predicts  just  such an   anti-correlation  between satellite   galaxy
abundance and  IHL  fraction.     Of   particular note  is   the  tight
correlation that emerges for group-scale objects when considering only
the   brightest of the survivors   within   the groups.  Our  analysis
indicates  that  for  galaxy-sized halos, the   68\%   scatter in each
$N_{surv}$ bin differs  by  roughly a  factor  of two  from the  bin's
median  value   and  is approximately constant    across  the range of
$N_{surv}$.  In group-scale  hosts, the variance  is generally smaller
($\sim{}\pm{}0.1-0.2$ in log$_{10}(N_{surv})$) and increases slowly as
the number of surviving massive galaxies grows.

This result may explain why 
some of the \citet{Gonzalez_etal07} clusters have higher
IHL fractions than we predict (e.g. Figure~\ref{fig:difftotal3}).  These
clusters were selected to have clearly dominant BHGs -- in other words,
to have a less dominant bright satellite population.  Based
on Figure~\ref{fig:nsurv}, we would expect
these systems to have higher IHL fractions than typical clusters of the
same mass.

\subsection{Tests for Robustness}
\label{sub:robustness}

To be sure, our model has several uncertain and poorly constrained 
elements.  Particular examples include the criterion for removing light from 
bound satellites and assigning it to the IHL, as well as the evolution of 
stellar mass with time.  Our argument in \S~\ref{subsec:exp} indicates 
that the overall trends for the IHL that we describe are set primarily 
by the convolution of the mass function of accreted subhalos with the 
mass-to-light ratios of accreted objects as a function of mass.  Further, 
because we are interested in the IHL {\em fraction} relative to 
the total luminosity of the system, errors in the normalization 
of the stellar mass function tend to offset each other, if not divide out precisely. 
These lines of reasoning suggest that the IHL 
trends that we outline should be robust, at least at the 
qualitative level, but likely at the quantitative level as well.  
Nevertheless, we have subjected our model to significant variations 
in parameter values to assess the robustness of the fiducial result.

Our free parameters govern
\begin{itemize}
\item Dark halo and galaxy disruption via $f_{crit}$ -- the
fraction of the initial halo circular velocity
that defines the critical circular velocity
below which satellite galaxies are
 deemed ``disrupted'' and their stellar mass is 
added to the diffuse light;
\item Star formation via $\alpha$  -- the star formation parameter
defined in Equation~(\ref{eq:sf});
\item Galaxy luminosities via $M/L_c$ as a function of mass -- 
the function set from large-scale galactic 
observations to relate central galaxy luminosities
and dark halo masses.
\end{itemize}

 \begin{figure}[b!]
 \plotone{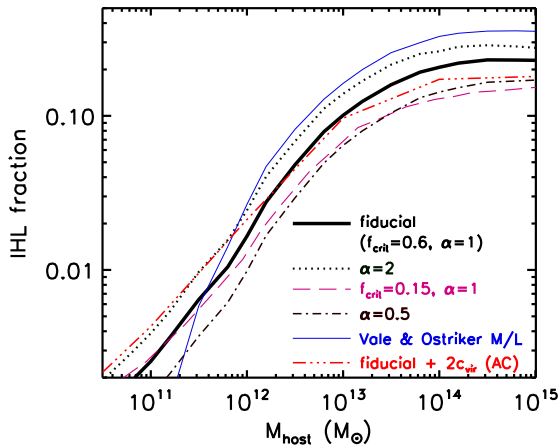}
\caption{An assessment of the robustness of our model for the IHL fraction (Case 1) 
under the variation in several inputs.  
The {\em thick, solid} line shows our fiducial result for the IHL fraction 
as a function of halo mass.  The {\em dotted} and {\em dash-dotted} lines 
show variations in our prescription for the 
evolution of stellar mass with time [Eq.~\ref{eq:sf}].  Specific parameter 
values are shown in the legend in the lower right corner of the plot.  
The {\em long-dashed} line shows a variation in 
the parameter $f_{crit}$ that describes the point at which the 
luminosity of a subhalo is assigned to the IHL.  The {\em dash-triple-dotted} 
line represents the Case 1 result using halo concentration parameters exactly 
twice their fiducial values, to imitate the effect of cooling baryons in the central 
regions of a halo.  Finally, the 
{\em thin, solid} line shows the result of using our fiducial model 
for the IHL along with the mass-to-light ratio of \citet{vale-2004-353} 
rather than that of \citet{yang_etal03}.  All models show the same gross 
features.  Quantitatively the model is very sensitive to the  
mass-to-light ratios of infalling objects in the range within which 
this input can be reliably constrained by independent means.
}
\label{fig:robustihl}
\end{figure}

In Figure~\ref{fig:robustihl} we plot the median IHL fraction computed
for our   fiducial model parameters  (thick  solid) along with various
other choices.  Note that for simplicity
we have neglected the ``pre-processed'' IHL
contribution in this set of tests.  We  find  that changing the  star  formation parameter
$\alpha{}$ over a very wide range ($ 0.5 < \alpha < 2$) 
produces global IHL trends that differ by less 
than a factor of   two from  the fiducial  case.
Predictably,  the   choice  of   $f_{crit}=0.15$  with  fiducial  star
formation ($\alpha=1$) results in less  diffuse light across the  full
mass  range because a dark matter subhalo is required to be more 
severely affected by the host potential before relinquishing its mass to the IHL.
However, even this drastic adjustment to $f_{crit}$ produces IHL values 
that are within a factor of two of the fiducial result.

The most visible change to our main result comes from revising 
our adopted $M/L_c$ from the Yang et al. (2003) inference to an alternative
form advocated by Vale \& Ostriker (2004) (see Figure 2).  
The Vale \& Ostriker (2004) $M/L_{c}$ 
relation has a steeper ``valley'' and, as could be expected from
our discussion in \S~\ref{subsec:exp}, gives rise to a steeper
$\fihl$ relation.  Even in this case, the overall increase in IHL is 
no greater than a factor of $\sim{}2$ at the cluster scale, while the steep 
faint-end slope of the Vale \& Ostriker $M/L_{c}$ relation suppresses the 
diffuse light in small galaxies to below fiducial levels.  Despite 
our limited knowledge of star formation, the overall trend appears robust.  
The sensitivity of the IHL fraction to the assumed mass-to-light ratios 
for infalling objects over the range within which the mass-to-light 
ratios can be reliably constrained suggests that the uncertainty in 
this ingredient is a fundamental limitation to the quantitative accuracy 
of any study based on this or similar approaches.  In particular, 
the IHL in group-sized systems relies on the precise
location of the $M\sim{}M_t$ trough in Figure~\ref{fig:masslum}.  
Alternatively, Figure~\ref{fig:robustihl} 
indicates that it may be possible in the future to constrain the
$M/L_{c}$ relation between small halos and central galaxies
by measuring the slope of the IHL fraction as a function of host halo mass, 
though more accurate theoretical methods would need to be employed in 
order to bring this goal to fruition.

We have managed, throughout this paper thus far, to avoid concerning 
ourselves too much with the baryon dynamics that play a significant role 
in the central regions of the halos that comprise our dark-matter model.  However, in 
the innermost kpc of a dark-matter halo, the mass density can be dominated 
by baryonic material, thereby increasing the concentration parameter typically 
assigned to an NFW density profile and making subhalos harder to disrupt.  
The response of the dark-matter profile to the presence of these 
cooling baryons is often modeled via 
adiabatic contraction (AC) \citep[see, \eg][]{Blumenthal86,gnedin2004}.  In cosmological simulations 
incorporating radiative cooling as well as star formation and supernova feedback, 
\citet{Rudd_etal07} find that the effects of baryon contraction can be approximated by increasing halo 
concentrations at fixed mass by a uniform factor, nearly constant over 
the mass regime of relevance and slightly smaller than a factor of two.  Since this adjustment 
applies to host halos as well as subhalos, the effect is two-fold: satellites become relatively more 
resistant to disruption, while their hosts have higher central densities and
stronger tidal fields as a result of this contraction.  
In order to test the robustness of our model to these competing phenomena, we double the initial concentration parameters 
of each host halo and subhalo.  The resultant IHL fraction, shown in Figure~\ref{fig:robustihl}, 
has a slightly shallower slope in the "break" between power-law $M/L$ regimes, although the 
galactic-scale diffuse light is slightly larger that of the fiducial Case 1, and 
the intracluster stellar mass decreases by less than a factor of two.  Overall, the effect is
less important than some of the other uncertainties we consider here.  
We interpret the relative changes from our fiducial model 
to be a consequence of the fact that the IHL in larger host halos is governed by the disruption of
subhalos that are relatively small in comparison to the host,  $\sim 10^{11.5} \Msun$.
These satellites will be more resistant to tidal disruption if they are contracted.  The IHL in
small hosts is set by the most massive subhalos that merge, and these systems are strongly affected by
dynamical friction.  The dynamical friction force in the host halo will be
 enhanced because of the contraction and this enhances massive satellite destruction probability.


In closing, we reiterate that IHL {\em fractions} will be naturally less susceptible
to fiducial normalizations, and  that our  intra-halo light 
predictions are driven  primarily by the shape  of the $M/L_c$ function
convolved with the accretion histories of Figure~\ref{fig:subfrac}.
We conclude that our  general prediction is robust; explicitly stated, 
that the IHL fraction should rise from a very small value 
$\lesssim 1\%$ in low-mass galaxies
to an appreciable fraction $\gtrsim 20\%$ in cluster-sized systems.

\section{Discussion}
\label{sec:discussion}

The main conclusions of our work may be summarized as follows:
\begin{itemize}
\item[1.] The IHL fraction in dark matter halos of mass $M$ is
expected to increase dramatically from $\sim{}0.5 \%$ to $\sim{}20 \%$ 
as we examine systems from the size of small spiral galaxies 
($M \sim{}10^{11} \Msun$) to galaxy groups ($M \sim{}10^{13} \Msun$).  
The IHL-mass relation becomes flatter at a value of
$\sim{}20\%$ for $M \gtrsim{}10^{13} \Msun$, increasing weakly thereafter to $\sim{}30\%$ 
for host halos of mass $M \sim{}10^{15} \Msun$.  While varying the
empirical mapping between halo mass and galaxy luminosity can
produce a slightly higher cluster IHL fraction, $\sim 40 \%$, 
the overall trends are very robust and  
are governed by the well-known  
fact that galaxy formation efficiency varies as a function of mass scale
while dark matter accretion processes are roughly self-similar.
Specifically, the subhalos that ``build'' galaxy halos have much lower
luminous mass fractions than the subhalos that build galaxy groups.  

\item[2.] The IHL component within galaxy halos is dominated by the disruption of satellites
of stellar mass $\sim{}10^{8.5} \Msun$ while the IHL component in clusters
is built from more massive stellar systems $\sim{}10^{11} \Msun$.  We expect that
more massive galaxies will therefore be surrounded by 
more metal rich stellar halos, as has been suggested by
 recent observations 
\citep[][although \citealt{ferguson07} disputes this claim]{2005ApJ...633..821M}.

\item[3.] The variation in IHL fraction from system to system at a fixed
halo mass is driven by
variations in  the  accretion history.  Systems   with fewer surviving
satellites tend to have  higher diffuse light fractions.  The  scatter
at fixed  mass is larger in  galaxy-sized halos because the light tends
to be dominated  by a  small  number  of  massive satellite  accretion
events.    As   indicated by   Figure~\ref{fig:nsurv},   the number of
surviving satellite  galaxies  in a group  is expected  to  negatively
correlate with that group's IHL  fraction, providing an  observational
expectation which future surveys may potentially address.  This phenomenon 
may also provide insight regarding the comparison of our results to observation, 
in which \citet{Gonzalez_etal07} finds a slightly higher IHL fraction than our model 
predicts for group-scale hosts, possibly due to a selection effect in which their sample 
systems are typically dominated by their central galaxies, with relatively few bright 
satellites and thus a systematically larger IHL value.
\end{itemize}

Current   observations place loose   constraints  on the diffuse light
fraction on every mass   scale.
By all 
indications, IHL accounts  for less than   a few percent of the  total
stellar mass in large galaxy-sized  host halos 
\citep[see][for discussions concerning the Galactic halo and that of M31, respectively]
{Siegel_etal02,Guhatakurta_etal05},
 while  the   diffuse
stellar components   of  cluster-sized hosts are   typically about one
order of magnitude higher
\citep{Mihos_etal05,Zibetti05,Krick_etal06,Gonzalez_etal07}.
A pronounced ``break'' in the diffuse light below the cluster scale is even
reported \citep{ciardullo2004}.  These results are
in general agreement with our expectations.

\begin{figure}[b!]
\plotone{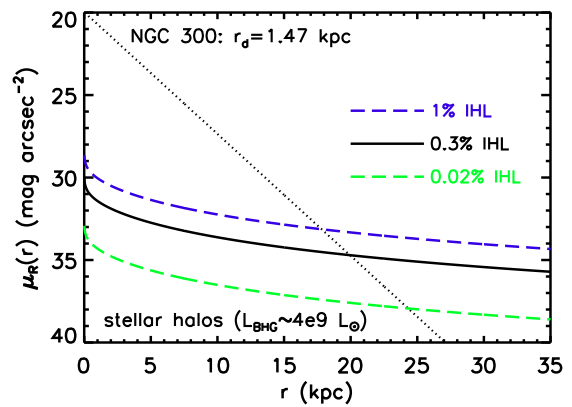}
\caption{The R-band surface brightness profile as a function of radius for the diffuse 
stellar component in a small host halo ($L_c \sim 4 \times 10^{9} \Lsun$), 
where the IHL is assumed to trace the background projected-NFW density profile.  Shown 
are the stellar halo profiles for the two values at either extreme of this host's IHL 95\% distribution, 
as well as the median value of $\fihl \simeq 0.003$. 
For comparison, we plot the surface brightness of the exponential disk of a similar 
system, the Sculptor group member NGC 300, a galaxy whose disk extends to at least 15 kpc without
the detection of an underlying diffuse component
\citep{hawthorn2005}. Our results demonstrate that this is not unexpected.    
}
\label{fig:surface}
\end{figure}

We predict that the   diffuse component around small  spiral  galaxies
will  contain  a very small   fraction  of the  primary galaxy's light on average, 
$\fihl \lesssim 1 \%$.  It is interesting to consider the surface brightness
limit that may be  required to observe  such a diffuse  component.  In
Figure~\ref{fig:surface} we investigate a simple example case where we
have distributed all of the diffuse light predicted for a low-luminosity
galaxy, $L_c \sim 4 \times 10^{9} L_{\odot}$, into an NFW halo density
profile that mirrors  that of the host halo.   Solid  and dashed lines
correspond to the median and 95 percentile predictions.  Here, we have
assumed   a stellar mass-to-light   ratio  of 1 in    the R-band.  For
reference we also plot the exponential  surface brightness profile 
\citep{kim2004} for the disk of a system of comparable luminosity, 
the Sculptor group galaxy NGC  300, which was  shown by
\citet{hawthorn2005} to extend $\sim 15$ kpc  from the galaxy's center
without revealing any  underlying diffuse component.  According to our
analysis,  a survey reaching $\sim  17$ kpc  from the galaxy's center 
and achieving 32  magnitudes per square arcsecond might
be  able  to detect a   stellar halo  around  NGC 300  if the  diffuse
component is comparatively bright, while a more  average IHL value for
the     system would require an even deeper search.
Similar analyses  for Milky-Way-sized stellar halos indicate
that  the  IHL  begins to separate   itself  from a (face-on) disk profile at
roughly  29-32 magnitudes  per square  arcsecond, which is in line with
the results of \cite{Irwin05} for M31.   This provides some
idea of the observational depth   that will, in the future,  be
required  to identify  remote    stellar  halos around small    spiral
galaxies.  It is worth pointing out that some fraction  of this  light may be  in the
form   of recently-destroyed  satellites, which should produce
 higher-surface brightness  features and  will  be more easily
seen \citep[\eg][]{bj05}.  Of course, the likelihood of a recent accretion will
decrease for lower mass galaxies.  A more detailed investigation of these
issues is warranted, but the complexity inherent in studying these 
issues places such an attempt beyond the scope of this paper.

It is worth noting that while we  have focused on accreted material as the source of diffuse light,
several other sources have been discussed.
These include  {\em in  situ} star
formation   \citep{gerhard2002},   ejection   from  binary     systems
\citep{holley-bockelmann2005}, dry mergers (in clusters) between ellipticals
\citep{stanghellini2006},   and  collisionless     evaporation
\citep{muccione2004}.   However, as this work demonstrates, galaxy disruption
provides a reasonable and seemingly inevitable 
mechanism for producing IHL on all scales, although these phenomena 
almost certainly have some subdominant role in the buildup of a host halo's diffuse light.

According to our picture, the driving force  behind the creation of IHL
on every  mass scale is the  stellar mass  spectrum of
intrahalo progenitors (Fig.~\ref{fig:dfdmstar}).  
The properties of   a system's diffuse luminous
component can  be understood   as the   result  of the
stochastically-driven  merger    history of stellar-rich     satellite
galaxies, indicating that future observations of intrahalo light could
be used  as a probe of a galaxy's merger history.
The  predicted  trend with IHL  fraction and  halo mass is
certainly  within the  scope of  future  observational work.
Interestingly, preliminary results from the
Galaxy Halos, Outer disks, Substructure, Thick disks, and Star clusters 
(GHOSTS) survey \citep{2007astro.ph..2168D} suggest that the stellar
halos of low-mass galaxies are, indeed, less prominent than those of
more massive galaxies.   Ongoing surveys of this type will be able to
test whether the expected trend carries over to other mass regimes.
As we demonstrated in Figure~\ref{fig:robustihl}, while the
qualitative positive trend between IHL fraction and halo mass is robust,
the {\em slope} of the relation is sensitive to the underlying
relationship between halo virial mass
and galaxy luminosity on dwarf galaxy scales.   
In principle, this measurement, in concert with models of the kind we present,
will help constrain the nature of galaxy formation in 
{\em dwarf-irregular-size halos} and test
the accretion histories of dark halos on small scales.

$\;$

We would like to thank Joel Berrier, 
Scott Chapman, Roelof de Jong, Annette Ferguson, Fabio Gastaldello,    Kathryn  Johnston, 
Jason Kalirai, Tammy Smecker-Hane,  Risa Wechsler, and Xiaohu Yang  for useful  discussions.  
Advice from Joss Bland-Hawthorn, John Feldmeier, Anthony Gonzalez, 
Andrey Kravtsov, and Dennis Zaritsky
greatly improved the paper and led to the creation of Figures 5, 6, and 10.  
We also thank the anonymous referee for suggestions that improved the quality of the paper.  
CWP and JSB are supported by National Science Foundation (NSF) 
grants AST-0607377 and AST-0507816, and the Center for 
Cosmology at UC Irvine.  ARZ is funded by 
the Kavli Institute for Cosmological Physics at 
The University of Chicago, by the NSF under grant PHY-0114422, 
and by the National Science Foundation Astronomy 
and Astrophysics Postdoctoral Fellowship program 
under grant AST-0602122.

\end{document}